\documentclass[12pt,a4paper]{article}
\usepackage{epsf,array,delarray,amsmath,stmaryrd,amssymb,amsthm}
\usepackage{graphicx}

\theoremstyle{plain}
\newtheorem{proposition}{Proposition}
\newtheorem{theorem}{Theorem}[section]

\newtheorem{definition}[theorem]{Definition}

\theoremstyle{definition}

\numberwithin{equation}{section}

\usepackage{cite}

\usepackage{graphicx}
\newcommand{\sA}{{\cal A}}
\newcommand{\sB}{{\cal B}}

\newcommand{\sG}{{\cal G}}

\newcommand{\sC}{{\cal C}}

\newcommand{\sY}{{\cal Y}}
\newcommand{\sI}{{\cal I}}

\newcommand{\sF}{{\cal F}}

\newcommand{\pp}{ {\mathbb P} }

\newcommand{\half}{  {\scriptstyle \frac{1}{2} } }

\newcommand{\tb}{\bar\theta}

\newcommand{\R}{\mathbb R}
\newcommand{\T}{\mathbb T}

\newcommand{\nl}{\vskip 0.05 in \noindent}

\newcommand{\bea}{\begin{eqnarray}}
\newcommand{\eea}{\end{eqnarray}}
\newcommand{\beast}{ \begin{eqnarray*} }
\newcommand{\eeast}{ \end{eqnarray*} }

\newcommand{\E}{E}
\newcommand{\be}{\begin{equation}}
\newcommand{\ee}{\end{equation}}
\newcommand{\ve}{\varepsilon}

\newcommand{\cL}{{\cal L}}
\newcommand{\bc}{{\bar{c}}}
\newcommand{\bC}{{\bar{C}}}
\newcommand{\bS}{{\bar{S}}}

\newcommand{\bt}{{\bar{\theta}}}
\newcommand{\bT}{{\bar{\Theta}}}
\newcommand{\bl}{{\bar{\lambda}}}

\newcommand{\tthe}{{\tilde{\theta}}}
\newcommand{\tThe}{{\tilde{\Theta}}}
\newcommand{\tl}{{\tilde{\lambda}}}
\newcommand{\tomeg}{{\tilde{\omega}}}
\newcommand{\tOmeg}{{\tilde{\Omega}}}
\newcommand{\tC}{\tilde C}
\newcommand{\tc}{\tilde c}

\newcommand{\tS}{\tilde S}

\newcommand{\tX}{\tilde X}

\begin{document}

\title{Diverse Beliefs}

\author{ A.A. Brown \thanks{Wilberforce Road, Cambridge CB3 0WB, UK (phone = +44 1223 337969 , email = A.A.Brown@statslab.cam.ac.uk)} \\ \small Statistical Laboratory, \\  \small University of Cambridge \and L.C.G. Rogers \thanks{Wilberforce Road, Cambridge CB3 0WB, UK (phone = +44 1223 766806, email = L.C.G.Rogers@statslab.cam.ac.uk)
We thank seminar participants at the Cambridge-Wharton meeting, June 2009,
particularly our discussant, Frank Diebold. 
\vskip 0.1 in
JEL Classifications:  D52, D53.
Keywords: Diverse beliefs, private information, probability, equilibrium,
beauty contest, mistaken beliefs.
}\\ \small Statistical Laboratory, \\ \small University of Cambridge }

\maketitle
\begin{center}
First version: July 2008
\end{center}
\vskip 0.2 in
\abstract{This paper presents a general framework for studying diverse
beliefs in dynamic economies. Within this general framework, the 
characterization of a central-planner general equilbrium turns out 
to be very easy to derive, and leads to a range of interesting 
applications. We show how for an economy with log investors holding
diverse beliefs, rational overconfidence is to be expected; 
volume-of-trade effects are effectively modelled; the Keynesian 
`beauty contest' can be modelled and analysed; and bubbles and
crashes arise naturally.
We remark that 
models where agents receive private information can  formally be
considered as models of diverse beliefs.}

\section{Introduction.}\label{intro}
Dynamic general equilibrium models provide us with perhaps
our best hope of understanding how markets and prices evolve,
but are  often frustratingly difficult to solve.  Representative
agent models are an exception, but the limitations of the representative
agent assumption are only too plain. Stepping up to models with 
many heterogeneous agents drastically reduces the available
range of tractable examples, but is a necessary approach to realism.
The simplest form of heterogeneity one could consider is one where
agents have different preferences, and perhaps different endowments,
but such models are not immediately suited to explaining effects arising
from different information, or from different beliefs, since the causes are
not being modelled.  

In a recent survey, Kurz \cite{Kurz} discusses the literature on models
with different information or beliefs, presents a compelling critique of 
models with private information, and expounds his own theory of how
to handle diverse beliefs. Models where agents receive private signals
about random quantities of interest have been extensively studied, 
but are in general hard to work with; see, for example, 
Lucas \cite{LucasJET},
Townsend \cite{Townsend},
Grossman \& Stiglitz \cite{GrossmanStiglitz},
Diamond \& Verrecchia \cite{DiamondVerrecchia},
Singleton \cite{Singleton},
Brown \& Jennings \cite{BrownJennings},
Grundy \& McNichol \cite{GrundyMcNichols},
Wang \cite{Wang}, He \& Wang \cite{HeWang},
Judd \& Bernardo \cite{JuddBernardo},
Morris \& Shin \cite{MorrisShin02}, \cite{MorrisShin05},
Hellwig \cite{Hellwig02}, \cite{Hellwig05}, 
Angeletos \& Pavan \cite{AngeletosPavan}.
 Problems
such as the Grossman-Stiglitz paradox, and the Milgrom-Stokey no-trade
theorem necessitate the introduction of exogenous noise into the models,
but nonetheless the treatment of private information is only tractable 
under very restricted modelling assumptions. There are also problems
at a conceptual level, as Kurz points out. Firstly, what is  private 
information?  In reality, the majority of agents' information is 
common, such as  macroeconomic indicators or 
the past performance of the stock, so we have to accept that a very 
small amount of private information might have a significant
impact. Secondly,
 if private information does exist, what could we say about it?  The 
private nature of the 
information would make it very difficult for us to verify any model that 
relied upon it.

For these reasons, we prefer to examine the class of models where all
agents have the same information, but interpret that information differently.
Although Kurz distinguishes such models from private information models, 
we can make the simple but important observation:
{\em a private
information model can be considered as a model where all agents have
common information, but have different beliefs about that information.}
Indeed, given a model where different agents receive private signals, we 
could regard this as a model where all agents receive the same information
but interpret it differently: every agent gets to see {\em
all } the private signals, but believes that the signals received by the others
are independent of everything else in the economy!  It would be hard to 
formulate a result general enough to cover all instances of this simple
principle, but in Section \ref{pidb} we state\footnote{
The proof is in Appendix \ref{PIDB}.
}  this in the
case of a finite-horizon Lucas tree model. This is our first main result,
 Theorem \ref{thmC}; its statement 
 is a little subtle, but illuminates the nature of the  equivalence.

In our treatment, the agents' different beliefs are modelled as different
{\em probability measures} $\pp^j$ defined over the same stochastic base
$(\Omega, \sF,(\sF_t)_{t \geq 0})$. Contrast this with the situation of private
information, where all agents share the same probability $\pp$, but work over
different stochastic bases $(\Omega, \sF^j,(\sF^j_t)_{t \geq 0})$.    The
diverse beliefs setting is far easier to work with, and, as we shall show, 
leads to simple but effective analyses.  The literature on diverse beliefs
is surveyed by Kurz, and includes the papers of 
Harrison \& Kreps \cite{HarrisonKreps},
Leland \cite{Leland},
Varian \cite{Varian85}, \cite{Varian89},
Harris \& Raviv \cite{HarrisRaviv},
Detemple \& Murthy \cite{DetempleMurthy},
Kandel \& Pearson \cite{KandelPearson},
Cabrales \& Hoshi \cite{CabralesHoshi},
Basak \cite{BasakJBF2005}, Basak \cite{BasakJEDC2000},
Basak \& Croitoru \cite{BasakCroitoru},
Calvet, Grandmont \& Lemaire \cite{CGL},
Wu \& Guo \cite{WuGuo03}, \cite{WuGuo04},
Buraschi \& Jiltsov \cite{BuraschiJiltsov}, Fan \cite{Fan},
Scheinkman \& Xiong \cite{ScheinkmanXiong},
Jouini \& Napp \cite{JouiniNapp},
Gallmeyer \& Hollifield \cite{GallmeyerHollifield},
Kogan, Ross, Wang \& Westerfield \cite{Koganetal}.
Among these, there are several which use the heterogeneity generated
by diverse beliefs to create interesting effects in a portfolio-constrained
setting; the papers  \cite{HarrisonKreps},
\cite{DetempleMurthy}, \cite{BasakJEDC2000},
\cite{BasakCroitoru},
\cite{ScheinkmanXiong},
 \cite{GallmeyerHollifield} are examples. The point is
that if (for example) short sales are constrained, the short-sales constraint
would never bind in an equilibrium with agents with homogeneous beliefs, 
because all would agree on the risk premium for all the assets, and if one
agent wanted to short a given asset, then so would all the others.  In all
these papers, as in \cite{BuraschiJiltsov} and \cite{DavidVeronesi}, there
is a model where agents have to filter a hidden process from observations
thereof, using different prior information; the analysis 
involves quite lengthy and detailed calculations based on some explicit
filtering problem formulation.  Our second main result, Theorem \ref{thm1},
is a general result, including all these examples, which characterizes the
equilibrium state-price density process in a complete
market, and hence all equilibrium prices. It is a result which does not
require much space to state, or to prove; {\sl how can it be so simple?}
Why are no lengthy calculations required? The answer of course is that
we are treating the equilibrium problem at a much greater level of 
generality than the papers cited earlier; we obtain a more general 
expression for the equilibrium, but for any particular example we would
need to specialize and calculate in order to derive   explicit solutions.
The essential element is to treat different beliefs as different equivalent
{\sl probability measures}; these are characterized by their likelihood-ratio
martingales, which are easy to work with.  

The characterization of the equilibrium given in Theorem \ref{thm1} is
not in essence new; in a single-period setting, Varian \cite{Varian85} 
expresses the first-order conditions in corresponding form; Jouini \& Napp
\cite{JouiniNapp} have the same characterization, which they then use to 
study a `market belief' (in the style of Calvet {\it et al.} \cite{CGL}).
However, it seems that
this simple and powerful characterization of the diverse beliefs
equilibrium has escaped general notice, so we hope to be forgiven for
emphasizing its importance, which we intend to demonstrate further
in the paper with several applications of the basic story.

In our account,  mutiple agents take positions in a single\footnote{
The restriction to a single asset is for notational convenience only; the 
entire analysis works also for multi-assets situations.
} 
asset which pays a continuous dividend stream, and is in unit net supply.
There is a riskless asset, in zero net supply. The agents have different
beliefs, represented as different probability measures, which we assume
with no loss of generality\footnote{
If agent $j$ has probability measure $\pp^j$, we could use the average
of the $\pp^j$ as a reference measure.
} are absolutely continuous with respect to some reference measure.
Though we have diverse beliefs, we stress that we do not take a continuum
of stochastically identical agents; agents' diversities do not just get replaced
by an average.  The form of the agents' beliefs is otherwise unrestricted:
\begin{itemize}
\item the agents could be stubborn bigots who assume they know the true 
distribution of the processes they observe and never change their
views;
\item the agents could be Bayesians updating their beliefs as time evolves;
\item the evolution of the agents' beliefs could be interlinked in various ways;
\end{itemize}
{\em all such 
structure is irrelevant at the first pass.}

Having derived the central-planner equilibrium in Section \ref{DBE}, we 
immediately show how this framework gives with no effort the result that
all agents are `rationally overconfident' - they all think that the particular
consumption stream that they have chosen is better than those chosen
by the others.

Obtaining explicitly-soluble examples with diverse beliefs is no easier
than in the situation where all beliefs are the same, and from Section \ref{Log}
onwards we make the simplifying assumption that all the agents have
log utilities.  This allows us to identify the state-price density process
quite explicitly, and to obtain expressions for the equilibrium price of the
asset, and for the riskless rate of return. We are also able to identify 
explicitly the portfolio of the risky asset over time for
each of the agents. In contrast to the common-beliefs situation, the 
portfolios have non-zero quadratic variation, which we interpret as a 
proxy for the volume of trade, and we study this  in  Section \ref{VoT}.

Section \ref{Beauty} addresses the `beauty contest' metaphor of 
Keynes  \cite{Keynes}. In 
this Section, we consider\footnote{
For this section only, we consider a simplified one-period example.
Keynes hypothesizes (but does not fully explain)
some mechanism which rewards beliefs which
are closer to the `average' belief. Our approach does not require us
to modify the objectives  of the agents.
} whether the individual agents in the model would
do better to {\em publicly profess beliefs they do not believe in.} The point
of doing this is that their objective is defined in terms of their true beliefs,
yet the equilibrium is characterized by the professed beliefs which guide
their investment decisions. It may be (and it turns out to be)
 that they may individually improve their objective by professing beliefs which
they do not hold. However, if all agents resort to this subterfuge, the only
Pareto-efficient solution results in lower welfare in some suitable sense.
The solution arrived at  is {\em a mixture of their true beliefs
and a population-average of beliefs.}
We contrast this with the recent 
study of Allen, Morris \& Shin \cite{AllenMorrisShin}, where the asset
prices are defined in terms of average expectation operators which 
do not compose in a time-consistent fashion. One consequence of this
is that the prices are not derived from a state-price density, whereas in 
our situation they are.  We believe that the time-inconsistency of their
average-expectation operators depends strictly on the overlapping-generations
structure assumed in their model, where each individual lives for 
just two periods. In such a story, an agent cannot {\em directly} compare
consumption now and consumption five periods in the future, because
five periods in the future he will not be consuming.  The comparison can 
only be via the intermediate pricing achieved in markets at the intervening
times. Indeed, in an overlapping-generations model with diverse beliefs
but with agents who live for a random length of time which may be 
arbitrarily large,
Brown \& Rogers \cite{BrownRogers} find a state-price density which
determines prices in the usual way.

In the next section, Section \ref{DMB},  we study the discrete-time
analogue of the continuous-time situation of Section \ref{Log}, but with
a difference. Starting from the observation that it is typically much easier 
to gather information on the stock price of a firm than on its dividend
process, we imagine now that some agents think that  {\em the stock price is a
multiple of the dividend } (as it would be in a homogeneous market.) 
Otherwise, they believe that the changes in the log dividend are
independent identically-distributed normal variables, whose variance
they know, but whose mean has a normal prior, which they attempt to
learn. Their beliefs are updated by the changes in {\em price}; but
their beliefs enter into the calculation of the price also, so there is a 
natural feedback mechanism from beliefs into prices. It is possible to 
carry the analysis quite a long way, but the story is ultimately too
complicated to study in general except by simulation. We present some
simulation results which show how the mistaken belief that the stock 
is a multiple of the dividend can produce some very substantial
and abrupt changes in price - bubbles and crashes.  In general 
terms, having more diligent\footnote{
We shall refer to an agent as diligent if he actually uses the changes
in log dividend - not the changes in log price - to update his beliefs.
} agents in the economy reduces the 
frequency and severity of these big changes.


We place in an appendix a very simple-minded model-fitting 
exercise; this is not because the study is not of intrinsic interest,
but rather because it differs in style from the mainly theoretical
body of the paper.
We take the diverse-beliefs  model with log agents and try to fit
 it to various 
sample moments of the dataset of Shiller\footnote{
This dataset can be downloaded from
{\tt http://www.econ.yale.edu/$\sim$shiller/data.htm}
}, as Kurz,
Jin \& Motolese \cite{KJM} do. We find good agreement using a model
with just three agents, and having reasonable parameter values. This
supports the view that diverse beliefs may be able to resolve the equity
premium puzzle, but the ability to match a few moments is not of course
sufficient to justify a statistical model.  Weizmann \cite{Weizmann},
 Jobert, Platania \& Rogers \cite{JPR}, Li \& Rogers
 \cite{LR_EPP}  analyze the equity premium
puzzle from the point of view of a representative Bayesian agent, and 
find reasonable values for parameter estimates, but do not present
evidence that the fitted models do any better than just fitting
constants to the data.

Section \ref{conclusions}
 concludes and maps out directions for future research.

\section{Equivalence of private-information and diverse-beliefs models.}\label{pidb}
 The purpose of this section is to show that any private-information (PI)
  equilibrium is (in a suitable sense) also a diverse-belief (DB) equilibrium. 
   It would be impossible to formulate a result broad enough to cover all
    imaginable instances of this principle, but what we shall do is to prove
     the result in the context of a discrete-time finite-horizon Lucas tree model
      with a single asset, and multiple agents.  To begin with there is a lengthy,
       necessary but straightforward statement of notation and definitions. 
       The main result is then expressed quite simply, but its rather lengthy
       proof is deferred to an appendix.
 
 The time index set is $\T=\{0,1,\dots, T\}$ for some positive integer $T$.  
 We suppose there is a single asset which delivers (random) output $\delta_t$
  at time $t\in \T$, and there is an $\R^d$-valued process $(X_t)_{t\in \T}$
   which we interpret as commonly-available information; we suppose that
    $\delta$ is one of the components of $X$.  There are $j$ agents, and in
     period $t$, agent $j$ receives private  signal $z^j_t$; we write $Z_t =
      (z^1_t,\dots, z^J_t)$ for the vector of all signals.   Agent $j$ has von
       Neumann-Morgenstern preferences over consumption streams 
$(c_t)_{t\in\T}$ given by
\be
E \left[ \sum^T_{t=0} U_j (t,c_t)\right] .
\label{vNM1}
\ee
The functions $U_j(t,\cdot):(0,\infty)\to \R$ are assumed concave, strictly
 increasing, $C^{2}$ and to satisfy the Inada conditions.  We write
  $\sG_t \equiv \sigma (X_s,Z_s : s \leq t)$ for the $\sigma$-field of all information
   at time $t$;  all filtrations considered will be sub-filtrations of $\sG$.

\begin{definition}
A {\em private-information equilibrium} with initial allocation  $y\in \R^J$ is a 
triple $(\bS_t, \bar{\Theta}_t, \bar{C}_t)_{t\in \T}$ of $\sG$-adapted processes,
 where $\bar{\Theta}_t = (\bar{\theta}^1_t,\dots, \bar{\theta}^J_t)$,
  $\bC_t=(\bc^1_t,\dots, \bc^J_t)$ and $\bS_t$ is real-valued, with the 
  following properties:
\begin{enumerate}
\item[(i)]
for all $j$, $\bc^j$ is adapted to the filtration $\bar{\sF}^j_t = \sigma
 (X_u, \bS_u, z^j_u : u\leq t)$ and $\bar{\theta}^j$ is previsible with
  respect to $\bar{\sF}^j$;
\item[(ii)] 
for all $j$ and for all $t \in \T$, the wealth equation 
\be
\bt^j_t (\bS_t + \delta_t) = \bt^j_{t+1} \bS_t + \bc^j_t
\label{wealthC}
\ee
holds, with the convention $\bS_T = \bt^j_{T+1}=0$;
\item[(iii)] 
for all $t \in \T$, markets clear:
\[
\sum_j \bt^j_t = 1, \qquad \sum_j \bc^j_t = \delta_t;
\]
\item[(iv)]
 $\bt^j_0 = y^j$ for all $j$;
\item[(v)] 
For all $j$, $(\bt^j, \bc^j)$ optimizes agent $j$'s objective \eqref{vNM1} over all
 choices $(\theta,c)$ of portfolio and consumption which satisfy the wealth
  equation \eqref{wealthC}, and such that $c$ is $\bar{\sF}^j$-adapted, 
  $\theta$ is   $\bar{\sF}^j$-previsible,  and $\theta_0= y^j$.
\end{enumerate}
\end{definition}

The notion of a PI equilibrium should be contrasted with the notion of
 diverse-belief equilibrium, where the filtration  is common, but the beliefs 
 are not.  So we shall suppose that there is some given filtration
  $(\sG_t)_{t\in\T}$ and some probability measure $P^j$ on 
  $(\Omega, \sG_T)$ for each $j=1,\dots, J$.  Agent $j$'s preferences 
  over consumption streams $(c_t)_{t\in \T}$ are given by 
\be
E^j \left[\sum^T_{t=0} U_j (t, c_t)\right]
\label{vNM2}
\ee
where $E^j$ denotes expectation with respect to $P^j$.

\begin{definition}
A {\em diverse-belief equilibrium} with initial allocation $y\in \R^J$ is a
 triple $(\tilde{S}_t, \tilde{\Theta}_t, \tilde{C}_T)_{t\in\T}$ of  
 $\sG$-adapted processes, where $\tThe_t = \left(\tilde{\theta}^1_t,\dots,
  \tilde{\theta}^J_t\right)$, $\tilde{C}_t = \left(\tilde{c}^1_t,\dots,
   \tilde{c}^J_t\right)$ and $\tilde{S}$ is real-valued, with
    the following properties.
\begin{enumerate}
\item[(i)] $\tThe$ is $\sG$-previsible;
\item[(ii)] for all $j$ and all $t \in \T$, the wealth equation
\be
\tilde{\theta}^j_t (\tilde{S}_t + \delta_t) = \tilde{\theta}^j_{t+1} \tilde{S}_t + \tilde{c}^j_t
\label{wealthC2}
\ee
with the convention $\tilde{S}_T = \tthe^j_{T+1}=0$; 
\item[(iii)] for all $t\in \T$, markets clear:
\[
\sum_j \tthe^j_t = 1, \quad \sum_j \tilde{c}^j_t = \delta_t;
\]

\item[(iv)] $\tthe^j_0 =y^j$ for all $j$;

\item[(v)] For all $j$, $(\tthe^j_0, \tilde{c}^j)$ optimizes agent $j$'s objective
\eqref{vNM2} over all choices $(\theta, c)$  of portfolio satisfying the wealth equation \eqref{wealthC2},
  and such that $c$ is $\sG$-adapted, $\theta$ is $\sG$-previsible, and
   $\theta_0=y^j$
\end{enumerate}
\end{definition}

\medskip
\noindent
Now that we have defined our terms, we are ready to state the main 
result.

\begin{theorem}\label{thmC}
Suppose that $(\bS, \bT, \bC)$ is a PI equilibrium with initial allocation
 $y\in \R^J$ for the discrete-time finite-horizon Lucas tree model introduced
  above.  Then it is possible to construct a filtered measurable space 
  $(\tOmeg, (\tilde{\sG}_t)_{t\in \T})$, carrying $\tilde{\sG}$-adapted 
  processes $\tilde X, \tS, \tThe,\tC$ of dimensions $d, 1, J$ and $J$ respectively, 
  and probability measures $P^j$, $j = 1,\dots, J$, on $(\tOmeg, \tilde{\sG}_T)$
   such that 
$(\tS_t, \tThe_t, \tC_t)_{t\in \T}$ is a DB equilibrium with initial allocation
 on $y\in \R^J$ and beliefs $(P^j)^J_{j=1}$ with the property that
$$
\cL (X, \bS, \bT, \bC) = \cL (\tilde X, \tS, \tThe, \tC).
$$
\end{theorem}

\noindent
{\sc Remark.}
Notice that the Theorem makes no statement about any analogue on
 the measurable space $(\tOmeg, \tilde{\sG}_T)$ of the signal process $Z$ 
 on the measurable space $(\Omega, \sG_T)$.  There may or may not be one. 
  Without compelling agents in the PI equilibrium to reveal these private  
  signals, the most it would be possible to observe would be the common
   knowledge $X$, the equilibrium price  $\bS$, the portfolio position $\bT$
    and the consumption choices $\bC$.  What the Theorem says is that
     the joint law of these processes (that is, the observables)  \emph{is} the
      joint law of the same observables in a DB equilibrium.  So from the point 
      of view of testing model predictions, there are no statistical properties 
      of a PI equilibrium which could not be explained by a DB equilibrium.  
      This justifies the claim that (for at least a finite-horizon Lucas tree model) we
       may ignore all (complicated) PI models and work only with (easier)
        DB models; \emph{PI equilibria are contained in DB equilibria}.
      The proof of Theorem \ref{thmC} is deferred to  Appendix \ref{PIDB}.

\section{Diverse beliefs equilibria.}\label{DBE}
We are going to derive a general equilibrium for a dynamic
economy with $J \geq 2$ agents, containing
 a single productive asset, whose output process
$(\delta_t)_{t \geq 0}$ is observable to all agents. We shall suppose
that time is continuous, and that $\delta$ is adapted to a filtration
$(\sF_t)_{t \geq 0}$ which is known to all agents. To cover various
technical issues, we shall assume that the filtered probability space
$(\Omega, \sF, (\sF_t)_{t \geq 0},\pp^0)$ satisfies the usual conditions;
see \cite{RogersWilliams} for definitions and further discussion.
For simplicity, we shall assume also that $\sF_0$ is trivial, so that
all $\sF_0$-measurable random variables are constant.

Though the $J$ agents all have the same information, they do not share
the same beliefs about the distributions of the processes they observe.
We suppose that agent $j$ thinks that the true probability is $\pp^j$, 
a measure locally equivalent   to $\pp^0$, with density process $\Lambda^j$
\begin{equation}
\Lambda^{j}_{t} = \frac{d \pp^j}{d \pp^0} \Bigl\vert_{\sF_t},
\label{Lambda}
\end{equation}
which is a positive martingale.

The objective of agent $j$ is to obtain
\begin{equation}
\sup \E^{j} \int_{0}^{\infty} U_{j}(t, c^{j}_{t}) \; dt
\end{equation}
where the supremum is over all  consumption policies which keep the 
wealth of agent $j$ positive.  Here, $U_{j}$ is some 
strictly increasing time-dependent utility, 
such that $U_{j}(t, \cdot)$ satisfies the Inada conditions.  
Notice that even if all agents have the same $U_j$, their objective
is calculated taking expectations under their different $\pp^j$, and so differences
in beliefs will result in different optimal behaviour.

The equilibrium for this market is determined in the following result.

\begin{theorem}\label{thm1}
Suppose that the market is complete\footnote{
The result holds also for a central-planner equilibrium; the essential
point is that there must be a common pricing of all contingent 
claims.
 In the case of a central-planner equilibrium, the constants
$\nu_j$ in the solution are determined by the weights on the 
individual agents.
}, and that integrability condition \eqref{IC} holds. 
 Then the unique equilibrium is 
determined by the state-price density process $\zeta$, which is related to
the individual agents' optimal consumption processes $c^j$ by
\begin{equation}
\boxed{
\nu_j \zeta_{t}   = U_{j}'(t, c_{t}^{j}) \Lambda_{t}^{j}
}
\label{SPD}
\end{equation}
for some constants $\nu_j>0$. The process $\zeta$ is determined from
the market-clearing condition and the $\nu_j$ by
\begin{equation}
\sum_{j} I_{j}(t, \zeta_{t} \nu_{j} / \Lambda_{t}^{j}) = \delta_{t},
\label{MC1}
\end{equation}
where $I_j$ is the inverse marginal utility $(U_j')^{-1}$ of agent $j$.
\end{theorem}

\medskip
\noindent
{\sc Proof.}
 Agent $j$'s objective can be written in the equivalent forms
\begin{equation}
\E^{j} \int_{0}^{\infty} U_{j}(t, c^{j}_{t}) \; dt
= \E^{0} \int_{0}^{\infty} \Lambda^j_t U_{j}(t, c^{j}_{t}) \; dt.
\label{obj2}
\end{equation}
Now consider the price that agent $j$ is willing to pay at time $s$ for a 
contingent claim which pays amount $Y_{t}$ at time $t>s$. Denote this 
price by $\pi_{s}^{j}(Y_{t})$ \footnote{Here, $Y_{t}$ is some bounded 
$\sF_{t}$-measurable random variable}.  By considering the change in 
agent $j$'s objective from buying this (marginal) contingent claim, the first 
order conditions give:
\begin{equation}
0 = \pi_{s}^{j}(Y_{t})\, U_{j}'(s, c_{s}^{j}) \,\Lambda_{s}^{j} - \E^{0} 
\left[Y_{t}\, U_{j}'(t, c_{t}^{j}) \, \Lambda_{t}^{j} \; |\; \sF_{s} \right] .
\nonumber
\end{equation}
Rearrangement gives
\begin{equation}
\pi_{s}^{j}(Y_{t}) = \E^{0} \left[ \; Y_{t}\,\frac{U_{j}'(t, c_{t}^{j})
 \Lambda_{t}^{j}}{U_{j}'(s, c_{s}^{j}) \Lambda_{s}^{j}}  
\;  \biggl\vert \; \sF_{s} \;  \right]
\label{pricing1}
\end{equation}
So we see that agent $j$ has state price density given by:
\begin{equation}
\zeta^{j}_{t} = U_{j}'(t, c_{t}^{j}) \Lambda_{t}^{j}
\end{equation}
As we assume that the market is complete,
 then the agents must agree on the
 price of all contingent claims. So looking at the expression for 
$ \pi_{s}^{j}(Y_{t})$ and recalling that $Y_{t}$ is arbitrary,
 we  must have
\begin{equation}
\zeta_{t,s}^{j} = \frac{U_{j}'(t, c_{t}^{j}) \Lambda_{t}^{j}}
{U_{j}'(s, c_{s}^{j}) \Lambda_{s}^{j}}
\nonumber
\end{equation}
is the same for all $j$. Hence 
\begin{align}
\nu_j \zeta_{t}  = U_{j}'(t, c_{t}^{j}) \,
 \Lambda_{t}^{j}  \nonumber
\end{align}
where $\nu_{j}$ is some $\sF_{s}$ random variable.  By taking $s=0$ and
invoking the triviality of $\sF_0$, we see that in fact $\nu_j$ must
be constant.

\nl
Now that we have \eqref{SPD}, deriving equilibrium prices follows 
from market clearing in the usual way. Defining\footnote{
The assumed properties of $U_j$ ensure that $I_j$ is well defined.
} the
inverse marginal utilities $I_j$ by
\begin{equation}
I_{j}(t, U_{j}'(t, y)) = y
\label{I_def}
\end{equation}
for any $y>0$, then 
\begin{equation}
I_{j}(t, \zeta_{t} \nu_{j} / \Lambda_{t}^{j}) = c_{t}^{j}.
\nonumber
\end{equation}
Summing on $j$ and using market clearing gives
\begin{equation}
\sum_{j} I_{j}(t, \zeta_{t} \nu_{j} / \Lambda_{t}^{j}) = \delta_{t}.
\label{MC1}
\end{equation}
This is an implicit equation for the unknown $\zeta$ in terms of the 
known quantities $\delta$ and $\Lambda^j$, and involving the 
constants $\nu_j$.
We shall assume the integrability condition:
\begin{equation}
\forall \; \nu_j>0, \quad  
 E^0 \biggl[ \int_0^\infty \zeta_t \delta_t \; dt \biggr] 
< \infty,
\label{IC}
\end{equation}
where $\zeta$ is determined from the $\nu_j$ by \eqref{MC1}.
The point of doing this is that  the stock price,
which is just the NPV of all future dividends, is given by
\begin{equation}
S_{t} = \E^{0} \left[\int_{t}^{\infty} \frac{ \zeta_{u} \delta_{u}}
{ \zeta_{t}} du \;  \biggl\vert  \; \sF_{t} \; \right] .
\label{stock}
\end{equation}
and we require that this be finite.

\hfill$\square$

\nl
{\sc Remarks.}
 (i) In the case where all agents have the same beliefs
(thus $\Lambda^J \equiv 1$ for all $j$), this reduces to the familiar 
expression for the state-price density as the marginal utility of optimal
consumption.
\nl
(ii) Notice that the situation is completely general; there is no assumption
about the nature of the stochastic processes, nor is there any assumption 
about the nature of the diverse beliefs. No such assumption is needed for
\eqref{SPD}.
\nl
(iii) {\bf Rational overconfidence.} 
Kurz remarks that ``a majority of people often expect to outperform 
the empirical frequency measured by the mean or median''.  In other 
words, each of the agents believes that they will usually do better than 
the average.  In our setup, this result comes for free.  If $\tilde{c}_{t}$ 
is any consumption stream and $c_{t}^{j}$ is agent $j$'s optimal 
consumption stream, then we have
\begin{equation}
\E^{j} \int_{0}^{\infty} U_{j}(t, c^{j}_{t}) dt 
\ge \E^{j} \int_{0}^{\infty} U_{j}(t, \tilde{c}_{t}) dt
\end{equation}
This follows simply from the fact that $c_{t}^{j}$ is agent $j$'s optimal
 consumption stream. In general, different agents will choose a different 
consumption stream, \emph{even if they have the same utility functions}; 
even if they do have the same utilities, each agent believes that he will do
 better (on average) than all the other agents.

\section{Log agents.}\label{Log}
Getting a reasonably explicit form for the state-price density process
$\zeta$ is key to making progress, and for the rest of the paper
unless explicitly stated to the contrary {\em  we shall make the
simplifying assumption}
\begin{equation}
\boxed{
U_{j}(t, x) = e^{-\rho_{j} t} \log x
}
\label{logU}
\end{equation}
for some positive $\rho_j$.  This leads to an explicit
form for  the state-price density, and from that,  expressions for
the wealth processes of the individual agents, 
the equilibrium price of the stock, and the equilibrium dynamics of the 
riskless rate when we assume specific dynamics for the dividend process.

\begin{theorem}\label{thm2}
With preferences given by \eqref{logU}, the state-price density process
is
\begin{equation}
\zeta_t = \delta_{t}^{-1} \sum_{j}\;  \frac{e^{-\rho_{j} t} 
\Lambda_{t}^{j} }{\nu_{j}}.
\label{LogSPD}
\end{equation}
The positive constants $\nu_j$ are fixed  in terms of the initial wealths
of the agents by 
\begin{equation}
w_{0}^{j} =    \Lambda_{0}^{j}/ \nu_{j} \rho_{j}
\label{Lognuw}
\end{equation}
where we make the convention that $\zeta_0=1$. At all times, the optimal
consumption rate processes are related to wealth by
\begin{equation}
    c^j_t = \rho_j\, w^j_t ,
    \label{cw}
\end{equation}
and the stock price is 
\begin{equation}
S_{t} = \delta_{t} \; \frac{ \sum_{j} e^{-\rho_{j} t} \Lambda_{t}^{j}/
\rho_{j} \nu_{j} }{\sum_{j} e^{-\rho_{j} t} \Lambda_{t}^{j} /\nu_{j}} .
\label{log_stock}
\end{equation}
\end{theorem}

\medskip
\noindent
{\sc Proof.}
Under the assumed form \eqref{logU} for the utility,
the relation \eqref{SPD} for the state-price density  simplifies to 
\begin{equation}
\frac{e^{-\rho_{j} t}  \Lambda_{t}^{j}}{c_{t}^{j}} = \nu_{j} \zeta_{t}.
\label{logSPD_1}
\end{equation}
The wealth process of agent $j$ is thus
\begin{eqnarray}
w_{t}^{j} & =& \E^{0} \left[\int_{t}^{\infty}
 \frac{ \zeta_{u} c_{u}^{j}}{ \zeta_{t}} \;du\;\biggl\vert\; \sF_{t}\right] 
\nonumber
 \\
& =& \E^{0} \left[\int_{t}^{\infty} 
\frac{ e^{- \rho_{j} u} \Lambda_{u}^{j} / \nu_{j} }
{ \zeta_{t}}\; du\;\biggl\vert\;  \sF_{t}\right] 
\nonumber
\\
& =& \zeta_{t}^{-1} e^{-\rho_{j} t} \Lambda_{t}^{j}/ \nu_{j} \rho_{j}
\label{wealth1}
\\
&=& c^j_t/\rho_j
\label{wealth2}
\end{eqnarray}
The derivation exploits the fact that $\Lambda^j$ is a $\pp^0$-martingale.
Using \eqref{logSPD_1}, market clearing gives
\begin{equation}
\delta_{t} = \sum_j c^j_t =\zeta_{t}^{-1} \sum_{j}\;  \frac{e^{-\rho_{j} t} 
\Lambda_{t}^{j} }{\nu_{j}},
\nonumber
\end{equation}
and hence by rearrangement
\begin{equation}
\zeta_t = \delta_{t}^{-1} \sum_{j}\;  \frac{e^{-\rho_{j} t} 
\Lambda_{t}^{j} }{\nu_{j}},
\nonumber
\end{equation}
which is \eqref{LogSPD}.
Since the stock is in unit net supply, and the bank account in 
zero net supply, we can quickly identify the stock price,
using \eqref{wealth1}:
\begin{equation}
S_t = \sum_j w^j_t =\zeta_{t}^{-1} \sum_{j}\; 
			\frac{e^{-\rho_{j} t} \Lambda_{t}^{j} }{\rho_j\nu_j}.
\label{S}
\end{equation}
Substituting from $\zeta$ from \eqref{LogSPD} leads to 
\begin{equation}
S_{t} = \delta_{t} \; \frac{ \sum_{j} e^{-\rho_{j} t} \Lambda_{t}^{j}/
\rho_{j} \nu_{j} }{\sum_{j} e^{-\rho_{j} t} \Lambda_{t}^{j} /\nu_{j}} 
\nonumber
\end{equation}

\hfill $\square$

\medskip
\noindent
Notice that in this case of log utilities, 
 the price-dividend ratio takes a particularly simple form:
\begin{equation}
\frac{S_{t}}{\delta_{t}} = \frac{ \sum_{j} e^{-\rho_{j} t} \Lambda_{t}^{j}
 /\rho_{j} \nu_{j}}{\sum_{j} e^{-\rho_{j} t} \Lambda_{t}^{j}/\nu_{j} },
\label{PDratio}
\end{equation}
which we shall have need of later when it comes to fitting various moments
to the Shiller dataset in Section \ref{EPP}. If all the agents have the same
beliefs, this is just a deterministic function of time, but with heterogeneous
beliefs this becomes a random process.  Notice also that the price-dividend
ratio depends only on the likelihood-ratio martingales, and not on the underlying
dividend process, though this property is special to the log case.

\vskip 0.1 in
This is about as far as we can get without some more specific assumptions
on the nature of the dividend process. The next result develops the equilibrium
under the assumption that the dividend process is an It\^o process.

\begin{theorem}\label{thm3}
Suppose that \eqref{logU} holds, and  that the dividend process
satisfies
\begin{equation}
d \delta_{t} = \delta_{t}  \sigma_{t} ( dX_{t} +\alpha^*_t \,dt )
\label{delta_def}
\end{equation}
where $X$ is an $(\sF_t)$-Brownian motion under $\pp^0$, and 
$\sigma$ is some positive bounded previsible process with bounded
inverse. Suppose that the agents' likelihood-ratio martingales $\Lambda^j$ obey
\begin{equation}
d \Lambda_{t}^{j} = \Lambda_{t}^{j} \alpha_{t}^{j} dX_{t}
\end{equation}
where the $\alpha^j$ are previsible processes\footnote{
Thus under the measure
$\pp^j$ the process $X$ becomes a Brownian motion with drift $\alpha^j$
(by the Cameron-Martin-Girsanov Theorem; see \cite{RogersWilliams}, 
IV.38 for an account). 
}.  Then the state-price
density process evolves as
\begin{equation}
d \zeta_{t} = \zeta_{t} ( - r_{t} dt - \kappa_{t} dX_{t})
\label{zetaBM}
\end{equation}
where
\begin{eqnarray}
 r_{t} & =&  \bar \rho_t + \sigma_t( \alpha^*_t + \bar\alpha_t) - \sigma^2_t,
\label{r}
\\
\kappa_{t} & =& \sigma_{t} -\bar\alpha_t.
\label{kappa}
\end{eqnarray}
The processes $\bar \alpha$ and $\bar\rho$ are weighted averages 
of the $\alpha^j$ and $\rho_j$:
\begin{equation}
 \bar \alpha_t \equiv \sum_j q^j_t \alpha^j_t, 
	\quad \bar \rho_t \equiv \sum_j q^j_t \rho_j, 
\label{alphabar}
\end{equation}
where 
\begin{equation}
	q_t^j \equiv  \frac{   e^{-\rho_{j} t} 
\Lambda_{t}^{j} /\nu_j   }
{
\sum_i  e^{-\rho_{i} t} 
\Lambda_{t}^{i} /\nu_i    
}  \,  .
\label{q_def}
\end{equation}
\end{theorem}

\medskip
\noindent
{\sc Proof.} 
The equation \eqref{LogSPD} for the state-price density gives
\begin{equation}
	\zeta_t \delta_{t}= \sum_{j}\;  \frac{e^{-\rho_{j} t} 
\Lambda_{t}^{j} }{\nu_{j}}\equiv L_t,
	\label{eq39}
\end{equation}
say.  A little It\^o calculus gives us
\begin{equation}
	dL_t = L_t ( \bar \alpha_t \, dX_t - \bar \rho_t \, dt)
	\label{eq310}
\end{equation}
where $\bar\alpha$ and $\bar\rho$ are as defined at \eqref{alphabar},
\eqref{q_def}
The dynamics of the riskless rate follow easily from
\eqref{eq39}, \eqref{eq310};  we have after some calculations that
\begin{equation}
d \zeta_{t} = \zeta_{t} ( - r_{t} dt - \kappa_{t} dX_{t})
\nonumber
\end{equation}
where
\begin{eqnarray}
 r_{t} & =&  \bar \rho_t + \sigma_t( \alpha^*_t + \bar\alpha_t) - \sigma^2_t,
\\
\kappa_{t} & =& \sigma_{t} -\bar\alpha_t,
\end{eqnarray}
as claimed at \eqref{r}, \eqref{kappa}.

\hfill $\square$

\medskip
\noindent
{\sc Remarks.}
(i) We can also derive the dynamics of the stock price. After some routine
calculations, we arrive at 
\begin{equation}
dS_t = S_t \bigl\lbrace (\kappa_t+ a_t) (dX_t + \kappa_t dt)
					+ r dt) \bigr\rbrace  - \delta_t dt,
\label{S_dynamics}
\end{equation}
where 
\begin{equation}
    a_t \equiv \frac{\sum \alpha^j_t e^{-\rho_j t} \Lambda_t^j / \nu_j\rho_j}
                  {\sum e^{-\rho_j t} \Lambda_t^j / \nu_j\rho_j}
\nonumber
\end{equation}
is an average of the $\alpha^j_t$ using weights different from the $q_t^j$.
This allows us to identify the volatility $\sigma^S$ of the equilibrium 
stock price, namely
\begin{equation}
   \sigma^S_t = \kappa_t + a_t = \sigma_t  - \bar\alpha_t + a_t.
\label{sigma_S}
\end{equation}
In general, this is different from the volatility $\sigma_t$ of the 
dividend process, {\em even if that volatility is constant}\footnote{
Compare with Kurz et al \cite{KJM} .}.
Observe also that if $\rho_j = \rho$ is the same for all $j$, then $a_t
=\bar\alpha_t$, and hence  $\sigma^S_t = \sigma_t$.
This checks out with what we would get from \eqref{PDratio}, which implies
that $\delta_t = \rho S_t$ when all the impatience parameters are the same.
\nl
(ii)
Notice also that
if all agents have the same beliefs, $\alpha^j_t = \alpha_t$ for
all $j$,  and $\alpha^*\equiv 0$, we see
\[
r_t = -\sigma^2_t + \frac{\sum_{j} e^{-\rho_{j} t}
 \rho_{j}/\nu_{j}}{\sum_{j} e^{-\rho_{j} t}/ \nu_{j}} +\sigma_t
 \bar\alpha_t;
\]
thus for constant $\alpha$ and $\sigma$, the riskless rate is a smooth 
deterministic function of time. By contrast,   if the $\alpha^j$ are 
constants but distinct, the agents have different
beliefs, and the riskless rate is truly stochastic.
\nl
(iii)  If all 
agents agreed, it is also immediate from \eqref{PDratio} that the 
volatility of the stock is the same as the volatility of the dividend
process; this illustrates
again the general principle that heterogeneous beliefs will generate
fluctuations which would be absent in a model where all agents agree.

\section{Volume of trade.}\label{VoT}
In Section \ref{Log} we derived the stock price process, and 
the individual wealth processes, when all agents had log utility.
This simple and explicit setup allows us to go further, and 
derive the portfolios held by the individual agents. This is of
interest because in the case where there is no diversity of 
belief, $\Lambda^j_t \equiv 1$ for all $j$, we see from \eqref{wealth1},
\eqref{S} that agent $j$'s wealth process $w^j_t$ is of the
form $w^j_t = g_j(t) S_t$ for some smooth deterministic function 
$g_j$. This implies that each agent's holding of the stock varies
smoothly and deterministically in time; in the extreme case where
all the $\rho_j$ are the same, there is no trade at all, and the 
agents simply stick with their initial holdings of the stock consuming
the dividend which it produces.  What we shall show in this section is
that even when all the agents have identical {\em time preferences}, that is,
all the $\rho_j$ are the same, diversity of belief generates a considerable
amount of trading, and (roughly speaking) the more diverse the beliefs are then the
more trading there is.  Compare with Harris \& Raviv \cite{HarrisRaviv},
and De Long {\it et al.} \cite{DeLong_etal}, who find that (in the context
of a private-information equilibrium) agent heterogeneity generates
trading.
Of course, one has to define what is meant
by volume of trading, since in the continuous-time setting  the
portfolio processes are typically of infinite-variation. We therefore
take as our definition of volume of trading the {\em quadratic variation of 
the agents' portfolios.}

\begin{theorem}\label{thm4}
Suppose that the assumptions of Theorem \ref{thm3} hold. With the notation 
of that Theorem, the number $\pi^j_t$ of units of the 
risky asset held by agent $j$ at time $t$ is
\begin{equation}
      \pi_{t}^{j} = \frac{ w_{t}^{j} ( \alpha_{t}^{j} +
 \kappa_{t} )}{ \sum_{i} w_{t}^{i} ( \alpha_{t}^{i} + \kappa_{t} )}
= \frac{w_{t}^{j} ( \alpha_{t}^{j} +
 \kappa_{t} )}{S_t (a_t+\kappa_t)}.
  \label{pi_t^j}
\end{equation}
Assuming further that $\sigma$ is constant, all $\alpha^j$ are constant,
and that $\rho_j = \rho$ for all $j$, the portfolio amounts $\pi^j$ have 
stochastic differential expansions
\begin{equation}
  d\pi^j_t = \theta^j_t dX_t +d \hbox{\rm( finite-variation terms)}
  \nonumber
\end{equation}
where
\begin{eqnarray}
     \theta^j_t &=& - \pi^j_t \bar\alpha_t + q^j_t  \bigl\lbrace\,
		\alpha^j( \sigma+\alpha^j - \bar\alpha_t) - v_t
\,\bigr\rbrace/\sigma
\nonumber
\\
    &=& q^j_t \biggl[ \; \frac{(\alpha^j-\bar\alpha)^2}{\sigma}
  - \frac{v_t}{\sigma}+\alpha^j - \bar\alpha   \;\biggr].
\label{thetaj}
\end{eqnarray}
\end{theorem}

\medskip\noindent
{\sc Proof.}
Starting from the  expression  \eqref{wealth1}
for the wealth,   an It\^{o} expansion gives
\begin{align}
dw_{t}^{j} = w_{t}^{j} \{- \rho_{j} dt + (\alpha_{t}^{j} +
 \kappa_{t} ) dX_{t} + (r_{t} + \kappa_{t}^{2} + \alpha_{t}^{j} \kappa_{t}) dt  \}.
\end{align}
However,  the wealth dynamics of agent $j$ can be expressed in terms of
the portfolio process $\pi^j$ as 
\begin{align}
dw_{t}^{j} = \pi_{t}^{j}( dS_{t} + \delta_{t} dt) - c_{t}^{j} dt
 + (w_{t}^{j} - \pi_{t}^{j} S_{t}) r_{t} dt.
\end{align}
Comparing coefficients  and using \eqref{S_dynamics} leads to the identification
\begin{align}
\pi_{t}^{j} = \frac{ w_{t}^{j} ( \alpha_{t}^{j} +
 \kappa_{t} )}{ \sum_{i} w_{t}^{i} ( \alpha_{t}^{i} + \kappa_{t} )}
= \frac{w_{t}^{j} ( \alpha_{t}^{j} +
 \kappa_{t} )}{S_t (a_t+\kappa_t)},
\nonumber
\end{align}
as asserted at \eqref{pi_t^j}\footnote{
 In the case where all the agents have the same belief, we have that:
\begin{align*}
\pi_{t}^{j} = \frac{ e^{-\rho_{j} t} / \nu_{j} \rho_{j}}{ \sum_{i} e^{-\rho_{i} t} / \nu_{i} \rho_{i}}
\end{align*} 
hence there is no volatility in the evolution of $\pi_{t}^{j}$.}

For the second part of the Theorem, we suppose that $\sigma$ is constant,
all the $\alpha_j$ are constant, and that $\rho_j = \rho$ for all $j$.  The 
expression \eqref{pi_t^j} for the proportion held by agent $j$ is now simply
\begin{eqnarray}
     \pi^j_t &=& \frac{w^j_t( \alpha^j + \sigma - \bar\alpha_t )}{\sigma S_t}
\nonumber
\\
	&=& \frac{\zeta_t w^j_t( \alpha^j + \sigma - \bar\alpha_t )}{\zeta_t\sigma S_t}
\nonumber
\\
	&=& \frac{\Lambda^j_t( \alpha^j + \sigma - \bar\alpha_t )}
                  {\sigma\nu_j (\sum \Lambda^i_t/\nu_i)}
\label{pij2}
\end{eqnarray}
The defining expression for $\bar\alpha_t$, simplified in this situation to
\begin{equation}
   \bar\alpha_t = \frac{\sum \alpha^j \Lambda^j_t/\nu_j}{\sum  \Lambda^j_t/\nu_j}
\; ,
\label{abar2}
\end{equation}
leads after some calculations to 
\begin{eqnarray*}
d\bar\alpha_t  &=& - \bar\alpha_t^2 \, dX_t 
           + \frac{\sum (\alpha^i)^2 \Lambda^i_t\nu_i}{\sum  \Lambda^j_t/\nu_j}
                     \;  dX_t + \hbox{\rm finite-variation terms}
\\
          &=& \frac{\sum (\alpha^i-\bar\alpha_t)^2 \Lambda^i_t/\nu_i}
                        {\sum  \Lambda^j_t/\nu_j}
                     \; dX_t + \hbox{\rm finite-variation terms}
\\
        &\equiv & v_t\, dX_t + \hbox{\rm finite-variation terms},
\end{eqnarray*}
say.
Suppose that $d\pi^j_t = \theta^j_t dX_t+$ finite-variation terms. Multiplying
\eqref{pij2} throughout by $\sum \Lambda^i_t/\nu_i$, and expanding gives
\begin{equation}
   \bigl\lbrace \, \theta^j_t + \pi^j_t \bar\alpha_t\, \bigr\rbrace
\Bigl( \sum \Lambda^i_t/\nu_i\Bigr)
	= \frac{\Lambda^j_t}{\sigma\nu_j}\, \bigl\lbrace\,
		\alpha^j( \sigma+\alpha^j - \bar\alpha_t) - v_t
\,\bigr\rbrace
\nonumber
\end{equation}
after some calculations. Rearranging, and recalling \eqref{q_def}, we 
obtain the expression
\begin{eqnarray*}
     \theta^j_t &=& - \pi^j_t \bar\alpha_t + q^j_t  \bigl\lbrace\,
		\alpha^j( \sigma+\alpha^j - \bar\alpha_t) - v_t
\,\bigr\rbrace/\sigma
\\
    &=& q^j_t \biggl[ \; \frac{(\alpha^j-\bar\alpha)^2}{\sigma}
  - \frac{v_t}{\sigma}+\alpha^j - \bar\alpha   \;\biggr]
\end{eqnarray*}
with some calculation, as asserted.

\hfill $\square$

\medskip\noindent
{\sc Remarks.} 
Notice that the sum of the $\theta^j$ is zero, as it must be, since the
sum of the $\pi^j$ is identically 1.
The absolute value of $\theta^j_t$ can be interpreted 
as the volume of trade in the risky stock by agent $j$.  Hence
the length of the vector $\theta$
 can be interpreted as the \emph{total} volume of trade. The
representation   \eqref{thetaj} shows that in general terms
the volume of trade gets bigger with greater diversity of 
beliefs, though it is hard to make this statement more precise.

\section{Diverse beliefs and beauty contests.}\label{Beauty}
The lively metaphor of a beauty contest, set forth by Keynes in 
Chapter 12 of his book {\sl The General Theory of Employment,
 Interest, and Money  } \cite{Keynes}, proposes a situation where
competitors have to pick the six prettiest faces from a set of 
one hundred photographs; the winner is the person whose chosen
six are the most chosen by all entrants to the competition.  By a 
rather  questionable extension of the metaphor, Keynes suggests
that the choice of a portfolio of stocks is rather like this,
where what matters is less the fundamental value of the stocks, 
but rather how the mass of market players perceive the values of 
the stocks. The metaphor has stuck in the popular imagination of
the subject, and leads to the idea that in some sense people should
adjust their beliefs according to what they think the population as
a whole believes.

In this Section, we substantiate this notion in a simple but
well-specified example.
We again use  the principle of modelling differences between agents
as differences in beliefs, but in a simpler setting, where there are
just two times, time 0 and time 1.
There is a finite set of $J$ agents, and a single
risky asset in zero net supply,  claims to which will be traded
at time 0, and whose random value $X$
will be revealed  at time 1.
Agent $j$ is a CARA agent, with utility $U_j(x) = - \gamma_j^{-1}
\exp( - \gamma_j x)$. The equilibrium analysis of this problem is 
of course very simple; we shall see what happens if agents are
allowed to pretend that their beliefs about the distribution of 
$X$ are different, and submit demands as a function of price
based on these fake beliefs. 

\begin{theorem}\label{thm5}
(i) If agent $j$ believes that $X \sim N(\alpha_j, v_j)$, then the
time-0 equilibrium price $S_0$ for a claim to a unit of the risky
asset at time 1 will be
\begin{equation}
   S_0 = \sum_j p_j \alpha_j,
\label{S0}
\end{equation}
where
\begin{equation}
   p_j \propto \frac{1}{\gamma_j v_j},  \qquad \sum p_j =1.
\end{equation}
\nl
(ii) If agents are allowed to pretend they have different beliefs, 
that $X \sim N(\tilde \alpha_j, v_j)$, then the  unique Pareto efficient 
choice is for agent $j$ to choose
\begin{equation}
    \tilde\alpha_j =   (1-p_j)\alpha_j + p_j \tilde S_0,
\label{tilde_alpha}
\end{equation}
where
\begin{equation}
 \tilde S_0 = \frac{\sum p_j(1-p_j)\alpha_j  }{\sum p_j(1-p_j)} .
\label{tilde_S0}
\end{equation}
The equilibrium price in this case is $\tilde S_0$.
\nl
(iii) In the equilibrium achieved in (ii), it is never the case that all 
agents improve their objectives, and it can be that all agents'
objectives are reduced.
\end{theorem}

\medskip
\noindent
{\sc Proof.}
(i) At time 0, the asset is on sale for (equilibrium) price $S_0$, and 
agent $j$ faces the optimization problem of choosing the number
$\theta$ of units of the asset to hold until time 1 with a view
to obtaining
\begin{equation}
	\max_\theta E_j [ \; U_j( \theta(X-S_0) ) \; ].
\nonumber
\end{equation}
 This is of course a simple calculation, resulting in the optimal 
portfolio choice
\begin{equation}
\theta = \theta_j \equiv \frac{\alpha_j - S_0}{\gamma_j v_j}.
\label{theta1}
\end{equation}
Market clearing now determines the equilibrium price $S_0$:
\begin{equation}
	S_0 = \sum_j p_j \alpha_j,
\label{S0}
\end{equation}
where $p_j \propto (\gamma_jv_j)^{-1}$, $\sum_j p_j = 1$. The agent's
maximized objective is 
\begin{equation}
     -\gamma_j^{-1}\exp\biggl(\;  -\frac{(\alpha_j-S_0)^2}{2v_j}
\; \biggr).
\label{obj1}
\end{equation}
\nl 
(ii) 
Given that agent $j$ is now professing to believe that 
$X\sim N(\tilde\alpha_j,v_j)$, the first analysis is repeated with 
tilded variables; we obtain equilibrium portfolios and price
\begin{equation}
 \tilde\theta_j =  \frac{\tilde\alpha_j - \tilde S_0}{\gamma_j v_j},
\quad\quad  
\tilde S_0 = \sum_j p_j \tilde\alpha_j.
\label{tilde_eqm}
\end{equation}
The objective of agent $j$ now becomes
\begin{equation}
  -\gamma_j^{-1}\exp\biggl(\; -\gamma_j\tilde\theta_j
(\alpha_j-\tilde\alpha_j)  -\frac{(\tilde\alpha_j-\tilde S_0)^2}{2v_j}
\; \biggr).
\label{obj3}
\end{equation}
If we consider the choice of agent $j$, assuming that the 
choices of all the other agents are given and fixed, then it is easy
to work \eqref{tilde_eqm}, \eqref{obj3} into the problem
\begin{equation}
   \max_{\tilde\alpha_j}\bigr\lbrace\;
(\tilde\alpha_j - \tilde S_0)(\alpha_j-\tilde\alpha_j)
+\half (\tilde\alpha_j-\tilde S_0)^2
\; \bigr\rbrace.
\label{obj4}
\end{equation}
A few lines of algebra lead to the conclusion that 
\begin{equation}
    \tilde\alpha_j =   (1-p_j)\alpha_j + p_j \tilde S_0.
\label{tilde_alpha}
\end{equation}
If we suppose that all agents have allowed themselves to profess
beliefs different from what they truly believe, then 
the relation \eqref{tilde_alpha}
must hold for each $j$; multiplying on both sides by $p_j$, summing
over $j$ and using \eqref{tilde_eqm} gives us
\[
   \tilde S_0 =  \sum_j p_j \tilde\alpha_j
= \sum p_j(1-p_j)\alpha_j + \sum p_j^2 \tilde S_0,
\]
which results in 
\begin{equation}
   \tilde S_0 = \frac{\sum p_j(1-p_j)\alpha_j  }{\sum p_j(1-p_j)} .
\label{tilde_S0}
\end{equation}
\nl 
(iii) Suppose the contrary: 
agent $j$ will do better in the Pareto efficient solution
 \eqref{tilde_alpha}, \eqref{tilde_S0} if and only if
\begin{equation}
(\alpha_j-S_0)^2 < (\tilde\alpha_j-\tilde S_0)^2 + 
	2(\tilde\alpha_j - \tilde S_0)(\alpha_j - \tilde\alpha_j);
\label{b3}
\end{equation}
see \eqref{obj1}, \eqref{obj4}. If we write $q_j = c p_j(1-p_j)$, 
where the positive constant $c$ is chosen to make $\sum q_j = 1$, 
we have \eqref{tilde_S0} that $\tilde S_0 = \sum q_j \alpha_j$, and so
\begin{equation}
   \sum q_j (\alpha_j - S_0)^2 = \sum_j q_j (\alpha_j - \tilde S_0)^2
			+ (S_0-\tilde S_0)^2.
\label{b4}
\end{equation}
However, the right-hand side of \eqref{b3} can be written as
\begin{equation}
   (\alpha_j-\tilde S_0)^2  - (\alpha_j - \tilde\alpha_j)^2\;  ;
\nonumber
\end{equation}
multiplying by $q_j$ and summing on $j$ gives us 
\begin{equation}
     \sum_j q_j (\alpha_j - \tilde S_0)^2  - \sum q_j (\alpha_j - \tilde\alpha_j)^2,
\label{b5}
\end{equation}
which is clearly less than \eqref{b4}, contradicting the supposition 
that \eqref{b3} holds for all $j$.

Numerical examples show that all agents' objectives may be reduced.

\hfill $\square$

\medskip
\noindent
{\sc Remarks.} (i)  
Notice that the  expression \eqref{tilde_S0} for the equilibrium price
$\tilde S_0$ is an average of the $\alpha_j$, as is the original
equilibrium price \eqref{S0}; but the weights are different.  The
interpretation of the expression \eqref{tilde_alpha} for 
$\tilde\alpha_j$ is that {\em the modified belief is the original belief
shifted a bit towards $\tilde S_0$,} rather in the style of Keynes.
However, the shifting of $\alpha_j$ is towards the {\em modified} average
$\tilde S_0$, not the original average $S_0$, and it is natural to 
ask whether in fact $\tilde \alpha_j$ lies between $\alpha_j$
and $S_0$. Numerical examples show that this is not always the
case, though it appears to be the majority case.  
\nl
(ii)
We have proved that \eqref{tilde_alpha}, \eqref{tilde_S0} is the
only possible Pareto-efficient modification of the beliefs of {\em all}
 the agents.
Could it be that there is some proper subset $F$ of $\{1,\ldots,J\}$
such that if the agents in $F$ profess different beliefs, and the agents
in $F^c$ do not, then no agent would wish to change their choice?
Modifying the preceding analysis, it is not hard to show that 
generically this does not happen; an agent who is given the choice
of whether or not to fake his beliefs, {\em ceteris paribus}, will always
want to do so.  This is not surprising, of course; given the freedom to 
optimize over a larger set, an agent would always prefer that.
\nl
(iii) Since the expression \eqref{b5} is less than \eqref{b4}, there
is not only a poorer objective for some agent, but also in some
collective sense the agents are doing worse.
\nl
(iv)
Notice that there are differences between the model we have 
studied here, and the Keynesian metaphor. In the latter, the payoff
is {\em entirely} determined by the behaviour of the population of agents,
whereas in our example, the random return $X$ is not affected by the 
acts of the agents, although the time-0 equilibrium price is. We find 
that, given the freedom to dissemble about their beliefs, agents will
do so, but that this will in general leave them worse off; it may even 
be that {\em all} are worse off as a result.

\section{Diverse mistaken beliefs}\label{DMB}
We have seen in Section \ref{Beauty} what happens in a simple
example where agents may dissemble about their
beliefs; the agents know what is going on, but they consciously act
differently.  In this Section, we shall study what is in some sense the
opposite situation, where the agents do not completely understand
the market around them, but nevertheless act in accordance with the 
analysis of Section \ref{Log}. We restrict the discussion
to agents with log utilities, and we shall work in discrete time\footnote{
The reason for this is that in continuous time the quadratic variation of
the observed stock process would not be consistent with the mistaken
beliefs which we propose to assign to some of the agents. 
}.

In practice, it may be very hard to learn about the dividends of an
asset; dividend payments are infrequent, and are often smoothed in 
various ways which limit their usefulness as indicators of the state of
a firm. On the other hand, the stock price is usually easy to get hold 
of; it is available daily or more frequently; and it provides what is 
arguably a more sensitive indicator of the state of the firm. In a 
market of log agents with common impatience
parameter $\rho$, the stock price is simply a multiple of the 
dividend process, $\delta_t = \rho S_t$; see \eqref{PDratio}. So we
shall consider a situation where {\em some agents observe the stock price,
and assume that it is a constant multiple of the dividend process.}
This introduces a natural and simple feedback mechanism from 
prices to beliefs. The agents assume that the log returns of the 
observed stock prices are actually the changes in $\log \delta$, and 
they modify their beliefs in the light of this knowledge - but those modified
beliefs then feed back into the stock prices.

To carry this analysis further, we record the following result, 
whose proof is a straightforward exercise.

\begin{proposition}
Suppose that $X_1, X_2,\ldots$ are independent $N(\mu, \tau^{-1})$ 
random variables, where $\tau$ is known, but $\mu$ is not known. Starting
with a $N(\hat\mu_0,(K_0\tau)^{-1}) $ prior for $\mu$, the posterior mean 
$\hat\mu_t$ for $\mu$, and the
posterior precision $\tau_t$ given $\sY_t \equiv \sigma( X_1,\ldots,X_t)$,
satisfy
\begin{eqnarray}
	\tau_t &=& K_t  \tau \equiv (t + K_0)\tau,
\label{tau_t}
\\
	K_t \hat\mu_t &=& K_0\hat\mu_0 + \sum_{i=1}^t X_i.
\label{muhat_t}
\end{eqnarray}
The joint density of $(X_1,\ldots,X_t)$ is 
\begin{equation}
  \lambda_t \equiv \exp\biggl\lbrace
	 - \frac{\tau}{2}\sum_1^t X_i^2 + \frac{\tau}{2}
	(K_t \hat\mu_t^2 - K_0\hat\mu_0^2) \biggr\rbrace \;
	\biggl(\frac{\tau}{2\pi}\biggr)^{t/2} \sqrt{\frac{K_0}{K_t}} .
\label{lambda_t}
\end{equation}
\end{proposition}

\noindent
{\sc Remarks.} (i) Notice that the joint density of $(X_1,\ldots,X_t)$
under the assumption that these are independent gaussians with
zero mean and variance $\tau^{-1}$ will be 
\begin{equation}
  \lambda^0_t \equiv \exp\biggl\lbrace
	 - \frac{\tau}{2}\sum_1^t X_i^2  \biggr\rbrace \;
	\biggl(\frac{\tau}{2\pi}\biggr)^{t/2} .
\nonumber
\end{equation}
Thus if we take this as the reference measure, the likelihood-ratio
martingale takes the simple form
\begin{equation}
  \Lambda_t = \lambda_t/\lambda_t^0 = \exp\biggl\lbrace \frac{\tau}{2}
	(K_t \hat\mu_t^2 - K_0\hat\mu_0^2) \biggr\rbrace \;
 \sqrt{\frac{K_0}{K_t}} .
\label{discrLa}
\end{equation}
\\
(ii) How does $\lambda_t$ change  to $\lambda_{t+1}$
when the new observation
$X_{t+1}$ is seen?  If we write
\begin{equation}
		X_{t+1} = \hat\mu_t + \ve,
\label{X(t+1)}
\end{equation}
then some simple calculations from \eqref{muhat_t} give us the
updating 
\begin{equation}
     \hat\mu_{t+1} = \hat\mu_t  + \frac{\ve}{K_{t+1}}.
\end{equation}
Using this and \eqref{lambda_t} we are able to derive the updating
\begin{equation}
   2\log(\lambda_{t+1}/\lambda_t) = -\tau\ve^2 \,\frac{K_t}{K_{t+1}}
+ \log\biggl( \frac{K_t}{K_{t+1}}\biggr)  + \log(\tau/2\pi)
\label{la_update}
\end{equation}
for $\lambda$.

\vskip 0.25 in
Working in discrete time, the arguments of Sections \ref{DBE} and
\ref{Log} go through with minor change, giving us 
\begin{equation}
\zeta_t \delta_t = \sum_j e^{-\rho_j t} \Lambda^j_t/\nu_j
\label{LogSPDdisc}
\end{equation}
exactly as before \eqref{LogSPD}, and the analogue
\begin{eqnarray}
   \zeta_t S_t &=& \sum_j \frac{e^{-\rho_j t} \Lambda^j_t}{\nu_j
(e^{\rho_j} - 1)}
\label{Sdisc}
\\
&\equiv & \sum_j \frac{e^{-\rho_j t} \Lambda^j_t}{\tilde \nu_j}
\end{eqnarray}
of \eqref{S} for the ex-dividend stock price $S_t$ at time $t$.

As we remarked earlier, the agents are supposed to see the
stock  price and assume that it is a multiple of the dividend
process. The discrete-time
 analogue of the dynamics \eqref{delta_def}
assumed previously for $\delta$ is to suppose that the random
variables $X_t \equiv \log(\delta_t/\delta_{t-1})$ are independent
$N(\mu,\tau^{-1})$. {\em Thus the agents will assume that the
random variables $\log(S_t/S_{t-1})$ are independent 
gaussians with common (unknown) mean and (known) precision\footnote{
We move to discrete time because in continuous time the 
quadratic variation of the price process would immediately tell
the agents that this hypothesis is false.
}.}
If we have determined the $\lambda^j_n$ and $S_n$ for 
$n \leq t$, we use the price/dividend ratio from \eqref{LogSPDdisc}
and \eqref{Sdisc} to determine the value of $\xi \equiv
\log(S_{t+1}/S_t)$:
\begin{eqnarray}
\frac{S_{t+1}}{\delta_{t+1}}&=& \frac{S_t e^\xi}{\delta_{t+1}}
\nonumber
\\
	&=& \frac{\sum_j e^{-\rho_j(t+1)} \lambda^j_{t+1}/\tilde\nu_j
}{\sum_j e^{-\rho_j(t+1)} \lambda^j_{t+1}/\nu_j}
\nonumber
\\ 
&=&\frac{\sum_j e^{-\rho_j(t+1)} (\lambda^j_{t+1}/\lambda^j_t)
\lambda^j_t/\tilde\nu_j
}{\sum_j e^{-\rho_j(t+1)}(\lambda^j_{t+1}/\lambda^j_t)
\lambda^j_t/\nu_j}.
\label{update}
\end{eqnarray}
In the expression \eqref{update}, everything is known except the ratios
$\lambda^j_{t+1}/\lambda^j_t$; and these are related 
(via \eqref{la_update} and \eqref{X(t+1)}) to the unknown value
$\xi = \log(S_{t+1}/S_t)$. Hence we are able to find (numerically)
the value of $\xi$ which solves the updating equation, and from this
work out how the price of the asset evolves. To make a meaningful
comparison, we consider the ratio of the price $S_t$ (which 
arises under the mistaken belief that the price is a multiple of the
dividend) to the price $S^*_t$ which arises if the agents are 
able to observe the dividend process exactly.  If this ratio is
close to one, then the effects of the mistaken assumption is 
small.

The combined effects of all these assumptions are too complicated
to be analyzed except numerically, so we have carried out a number
of simulations. Throughout, we supposed that the annualised volatility
of the dividend process is $0.25$, the actual annualised growth
rate is $1.5\%$, and the time between observations is one day
(thus the moments of each log price change are those implied by
the annualised figures). 

The characteristics of the agents are generated randomly.
One feature which we took care to build in is that if we perform a
simulation with $n_1$ agents, and then repeat with the same
random seed but with $n_2>n_1$ agents, then the first $n_1$
agents in the second simulation are identical to the $n_1$ agents
used in the first. The distributions of the different characteristics
are as follows. The $\rho_j$ are supposed to be 
drawn uniformly from  $[0.04,0.33]$, corresponding to 
mean look-ahead times ranging from 3 to 25 years. 
The assumed values of $\tau$ for the agents are drawn 
uniformly from $[0.4\tau^*, 1.05\tau^*]$, where $\tau^*$
is the true value used for the simulations. The prior means
for the annualised growth rate were drawn uniformly from
$[-0.05,0.15]$, and all the $\nu_j$ are assumed to be 
equal to 1.

We performed a number of runs with the same random 
seed (and therefore the same realised sample path of $\delta$)
for 30 agents, and for 50 agents. The different runs were also 
distinguished by the different numbers of agents who are
assumed to be diligent, in the sense that some of the agents
might update their posteriors {\em seeing the true values of
$\delta$, and believing that they are correct.}  Thus if all
the agents were diligent, then the prices observed are formed
exactly as described in Section \ref{Log}; the ratio of the
ideal stock price $S$ to the dividend process is given by 
\eqref{PDratio}.We denote this ideal stock price by
$S^*$ for the purposes of the discussion of this section,
to distinguish it from the price $S$ actually computed at
\eqref{update}. The various figures shown come as two 
panels, the upper showing the log of the ratio $S^*/\delta$,
and the lower showing the log of the ratio $S/S^*$. The different
figures differ in the number of agents assumed to be diligent;
for the same total number of agents, the upper panel should be
the same, and visual inspection shows that this is the case. 

\begin{figure}[H]
\begin{center}
\includegraphics[height=16.5cm,width=14.5cm,angle=270]
{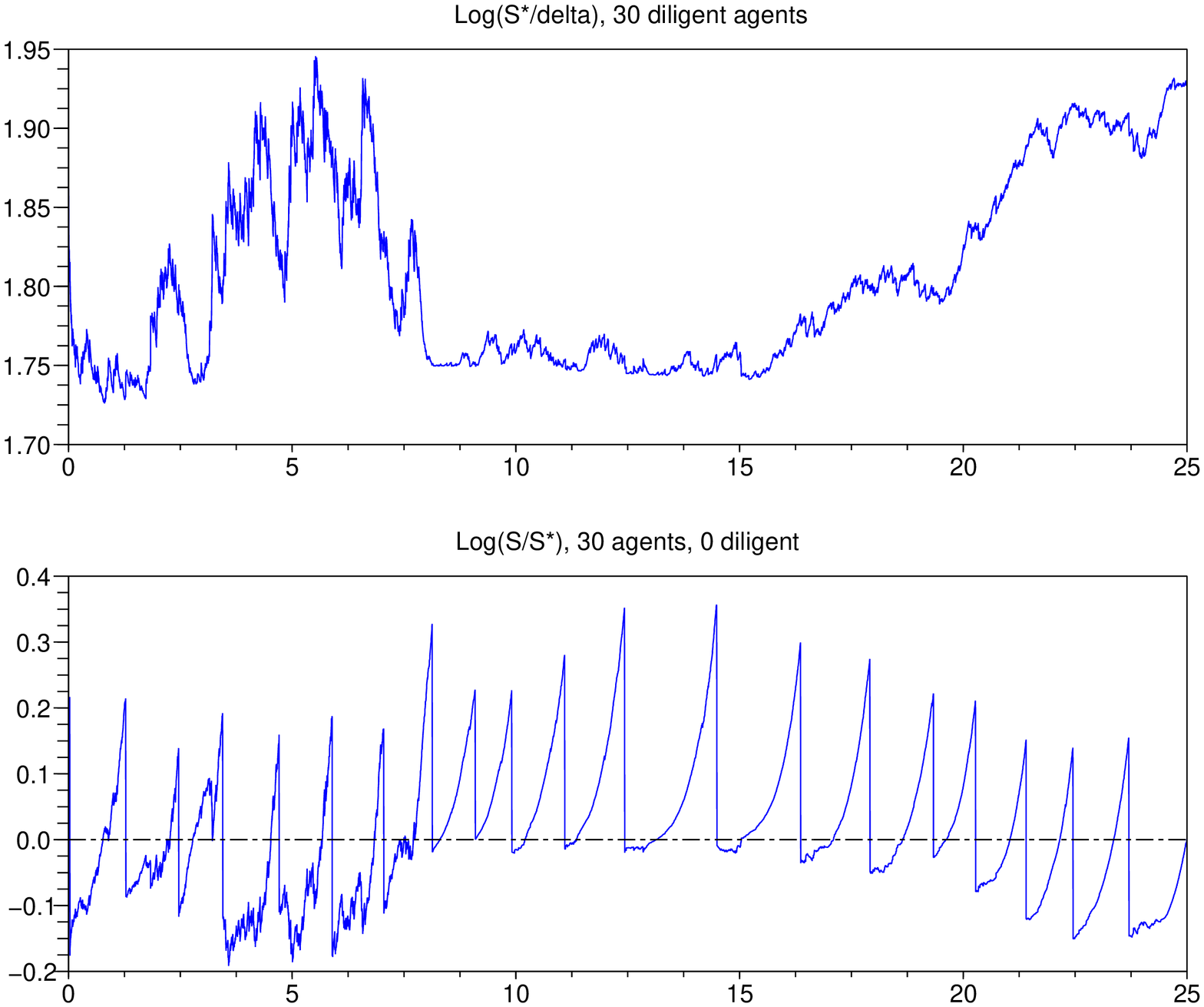}
\end{center}
\caption{
}
\label{Fig5_25Y_30A_0K}
\end{figure}

For 30 agents, we show in Figure \ref{Fig5_25Y_30A_0K} the
 behaviour of the price
when no agent is diligent; the repeated ramping up followed
by sharp falls is the most obvious feature\footnote{
Other simulations generate ramping down followed by sharp
rises.
}, and the range of 
values covered is quite high, from about -0.2 to nearly 0.4.
Changing one agent to diligent, we still see a choppy price path,
Figure \ref{Fig5_25Y_30A_5K},
though the ramp-ups are less pronounced, and the overall range of
the trajectory is smaller. The overall level however is quite different.
Increasing the number of diligent agents to 5,
Figure \ref{Fig5_25Y_30A_5K} largely eliminates the
peaky behaviour of the previous two plots, and it would be natural 
to conjecture that this more orderly behaviour becomes more prevalent
as the number of diligent agents rises, but the plot  Figure
\ref{Fig5_25Y_30A_10K} with
10 diligent agents suggests otherwise. The final plot 
Figure \ref{Fig5_25Y_30A_25K} in the
series, with 25 diligent agents, still shows quite a wide range of variation
of $S$ from $S^*$; only one in six of the agents is mistakenly
interpreting the price as a multiple of the dividend, and yet the 
log of the price ratio ranges from below -0.2 to over 0.1.

The Figures  \ref{Fig5_25Y_50A_0K}, 
 \ref{Fig5_25Y_50A_1K},  \ref{Fig5_25Y_50A_5K},  \ref{Fig5_25Y_50A_10K},
 \ref{Fig5_25Y_50A_25K},  \ref{Fig5_25Y_50A_35K},
  show the corresponding results for 50 agents,
with similar qualitative features; notice particularly the dramatic
crash when no agent is diligent!

\vskip 0.15in
What we see in the simulations are qualitative features
of bubbles and crashes, 
which one might also try to explain by models of  herding, 
or   behavioural effects. However, it is not necessary to 
construct such models to exhibit these phenomena.
The present framework is able to 
generate such qualitative features strictly within the neoclassical
framework of finance; all agents are behaving rationally, 
the only point is that they have misinterpreted what the 
market prices actually are. This  market
is definitely {\em not} always right.

\section{Conclusions}\label{conclusions}
This paper has shown how to deal with diverse beliefs of agents in a 
completely general manner; the key observation is that we should model
agents' beliefs as probability measures, whose likelihood-ratio martingales
enter naturally into the optimality criterion, and thence into equilibrium 
prices.  

The first consequence of this approach is that we are able to 
show that\footnote{
.. in the context of a finite-horizon Lucas tree model ...
}  equilibria where agents have diverse private information
are indistinguishable from equilibria where agents have common
information, but different beliefs.  This allows us to restrict attention
to (analytically simpler) diverse-beliefs equilibria.

Abstract expressions for the state-price density and for the 
equilibrium stock price arise simply from the assumptions, and are visibly
analogous to (but extensions of) the corresponding expressions with 
no diversity of belief. An immediate first result is an explanation of
 the phenomenon of rational overconfidence.

By specializing to the case of log agents, the equilibrium can be
computed quite explicitly, and its properties studied. We find quite
simple and explicit expressions for the riskless rate, the stock price,
the risk premium and the volatility of the stock price, in terms of the 
fundamentals of the problem, namely, the dynamics of the dividend
process and the beliefs of the agents, expressed as likelihood-ratio
martingales. Diversity of belief generates an active market,
and we are able to find an expression for the volatility of the agents'
holdings of the stock, which we interpret as a proxy for volume of 
trade. In general, greater diversity of belief generates a larger volume of trade.

In a one-period example, we are able to show that under the assumption
of diverse beliefs, there is benefit to individual agents to act as if their
beliefs were different from what they truly believe; such actions modify
the equilibrium in such a way that there is loss of welfare,
but no one agent would change. The beliefs adopted are the original
beliefs shifted towards a population average.
 This is therefore an analysis which explains the 
`beauty contest' phenomenon commented on and postulated
by Keynes, using no modelling elements other than rational
expectations equilibrium and diverse beliefs. In particular, it is not
necessary to introduce any `behavioural' concepts, nor are the 
agents' objectives in any way unconventional.  

Staying within this strictly neoclassical financial framework, we
find a mechanism to generate bubbles and crashes, by supposing
that some agents assume that the observed stock prices are actually
constant multiples of the dividend process (as would be the case
in a homogeneous market).  Again, there is no need to use concepts from
behavioural finance - the bubble is generated by entirely rational
agents, some of whom happen to be rational and mistaken.

There remain many interesting questions to be studied in this
area. For example, can diverse beliefs create an economic
r\^ole for money, by (say) imposing leverage constraints which more
money will ease? The paper \cite{LiRogers} is a first step down this 
road. Are there tractable examples where the agents have utilities
different from log, and if so, what do the solutions look like\footnote{
An interesting extension of the log agent is \cite{AAB}, where agents
are supposed to be CRRA with an integer coefficient of relative risk
aversion.
}? These
and other questions are in principle amenable to a correctly-formulated
modelling of diverse beliefs, which this paper has attempted to present.

\pagebreak
\appendix

\section{Fitting annual return and consumption data.}\label{EPP}
Kurz \cite{Kurz} uses his model of diverse beliefs to fit various
sample moments of the Shiller data set, and we perform a similar
study here. 

 We take a very simple version of the model,
with just three agents who never change their beliefs, so we assume that the
 $\alpha^{j}$ are constant. We also take $\sigma_{t}$ to be constant.

 The quantities of interest are shown in the table below; we 
list both the empirical value\footnote{These empirical values are calculated 
by Kurz and are based on the Shiller data set. They are based on monthly data
 from the S\&P 500 between 1871 and 1998. See \cite{Kurz} and \cite{KJM} for
 further details.} and the values as produced by  fitting our model.

\begin{table}[h]
\caption{Simulation Results}
\begin{center}
   \begin{tabular}{ |l || c | c |}
    \hline
 & Fitted  & Empirical \\ \hline
Mean price/dividend ratio  & 26.06	& 25 \\ \hline
Standard deviation of price/dividend ratio  & 3.84  &  7.1  \\ \hline
Mean return on equity  & 0.077 &  0.07 \\ \hline
Standard deviation of return on equity  & 0.134  &  0.18 \\ \hline
Mean riskless rate  &  0.018  &  0.018 \\ \hline
Standard deviation of riskless rate  & 0.061  &  0.057 \\ \hline
Equity Premium  &  0.059 &   0.06 \\ \hline
Sharpe Ratio & 0.326  &   0.33 \\
    \hline
  \end{tabular}
\end{center}
  \label{Table1}
\end{table}
   
The results shown were generated by choosing $\sigma=0.517, 
\alpha^{*}=- 0.01, \alpha^{1}= 0.210, \alpha^{2}=0.727,
\alpha^3 = -0.05,
 \rho_{1}=0.131, \rho_{2}=0.01, \rho_3 = 0.443,
\nu_{1}=14.47, \nu_{2}=1.00, \nu_3 = 0.174$.

From the table above, we see that the diverse beliefs model with
these parameter values gives quite a good fit to the sample moments
considered by Kurz {\it et al.}.  Only the standard deviation of the 
price/dividend ratio
is substantially off the empirical value, a sample moment which we note
was not fitted very closely by Kurz either, probably because the
volatility of recorded annual consumption is in general too small to 
explain the observed volatility in stock returns. Nevertheless, the
model seems to be doing a reasonable job explaining these figures
given the very specific assumptions made.

\section{ Bayesian learning.}

The case in which all the $\alpha$ are constant corresponds to that in 
which the agents all start with a belief about the behaviour of the dividend
 process and stick with this forever.  Such a setup is in some senses
 unsatisfactory, because even if the agents were to observe that the
 behaviour of the dividend were very different to their initial beliefs about it,
 they would still keep with these initial beliefs.

We therefore consider the case of Bayesian agents, who learn as they 
observe data.  Bayesian learning is a huge topic which has been studied
by \cite{BBK}, \cite{BX},  \cite{GuidolinTimmermann},
\cite{DavidVeronesi}, \cite{relaxed}
among others.
For example,
 Guidolin and Timmermann \cite{GuidolinTimmermann}  look at a 
discrete time case in which the dividend process can have one of two
 different growth rates over each time period and the probability of each
 growth rate is unknown to the agents.  The agents are learning, so this
 affects the way that the stock price is calculated and hence the dynamics 
of the stock and options prices. Again, David and Veronesi
 \cite{DavidVeronesi} look at a continuous time model in which at any given
 time, the economy can be in one of two states; boom and recession. 
 The agents do not observe this state directly, but instead must infer it
 from their observations of the dividend process.  

We take a very  unsophisticated model of Bayesian learning, which
for completeness summarises a story told before else where;
see, for example, Brown, Bawa \& Klein \cite{BBK}, Brennan \& Xia
\cite{BX}, or Rogers \cite{relaxed} for much the same material.

An agent observes a Brownian motion with drift:
\begin{align*}
Y_t = X_t + bt
\end{align*}
where $X$ is  a $\pp$-Brownian motion and $b$ is some  unknown constant.
 Instead of making an initial guess at the value of $b$ and
 sticking with it,  the agent gives a prior distribution to the unknown
 parameter $b$ and then updates this prior distribution as time progresses. 
 If the agent was sure about $b$, then he  would have:
\begin{align*}
\frac{d\pp}{d\pp^0}\biggr\vert_{\sF_t}
\equiv \Lambda_{t} = \exp \{ b X_{t} - \frac{1}{2} b^{2} t \}
\end{align*}
However, the agent gives $b$ a normal prior distribution
with mean $\beta$ and precision $\epsilon$.
\footnote{This is equivalent to having variance $\epsilon^{-1}$} 
 It follows that the change of measure the agent works with is given by:
\begin{align*}
\Lambda_{t} &= \int_{-\infty}^{\infty} \sqrt{\frac{\epsilon}{2 \pi}}
 \exp \{- \frac{\epsilon}{2} (b' - \beta)^{2} + b' X_{t} - \frac{1}{2}
 (b')^{2} t  \}db' \\
& = \sqrt{\frac{\epsilon}{\epsilon + t}}  \;\;
 \exp \biggl\lbrace  \frac{X_{t}^{2} +
 2 \beta \epsilon X_{t} - \epsilon (\beta)^{2} t }
{2( \epsilon + t)} \biggr\rbrace
\end{align*}
This gives:
\begin{align*}
\Lambda_{t} = \Lambda_{t} \alpha_{t} dX_{t}
\end{align*}
where:
\begin{align}
\alpha_{t} = \frac{X_{t} + \beta \epsilon}{ \epsilon + t}
\end{align}

This is of the form described in Section \ref{DBE}, but the $\alpha_{t}$
 are now adapted processes rather than constants.  Thus, our model can deal
 with intelligent agents who update their beliefs, as well as the simple agents
 who always hold the same beliefs.

\begin{figure}[H]
\begin{center}
\includegraphics[height=16.5cm,width=14.5cm,angle=270]
{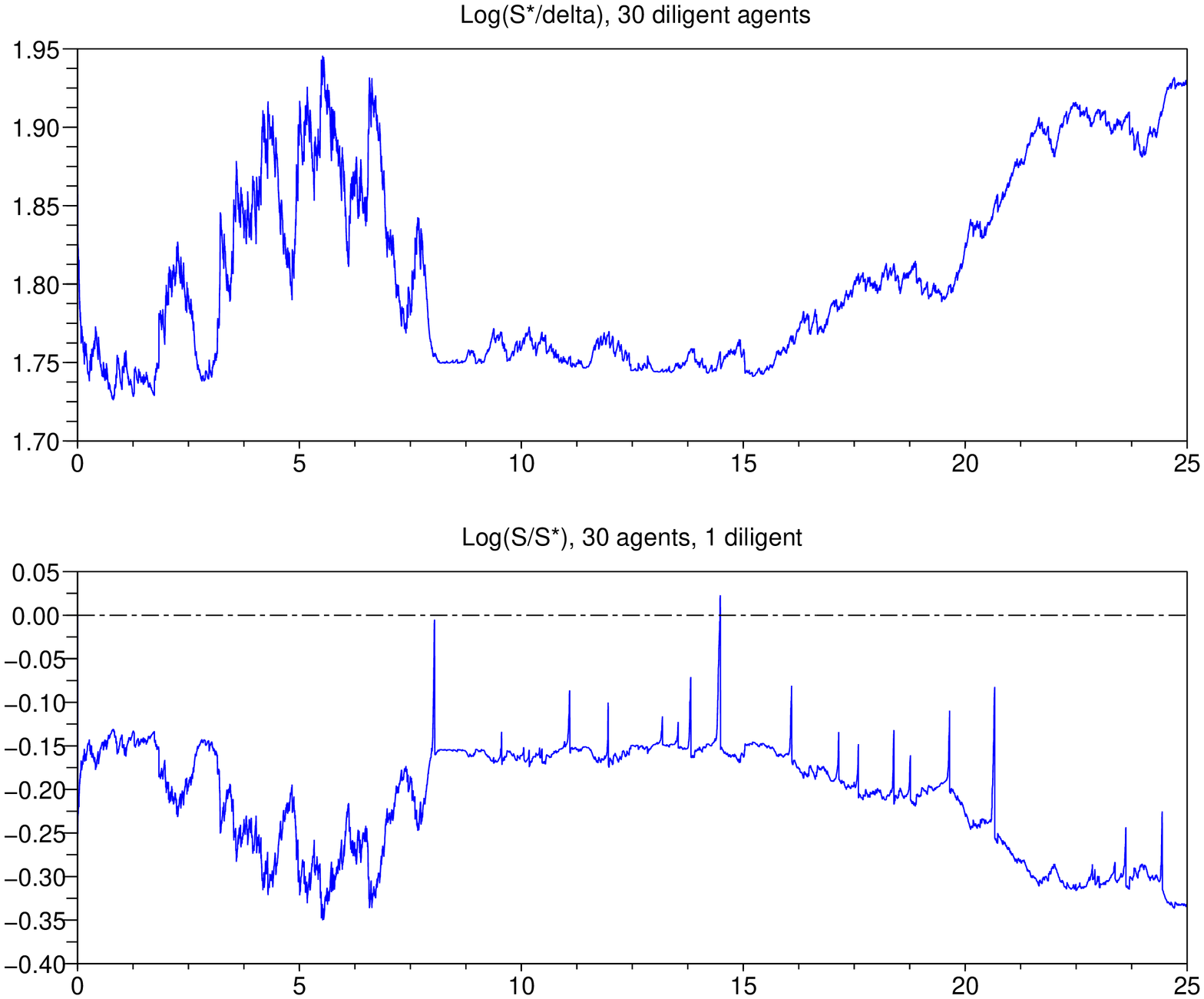}
\end{center}
\caption{
}
\label{Fig5_25Y_30A_1K}
\end{figure}

\begin{figure}[H]
\begin{center}
\includegraphics[height=16.5cm,width=14.5cm,angle=270]
{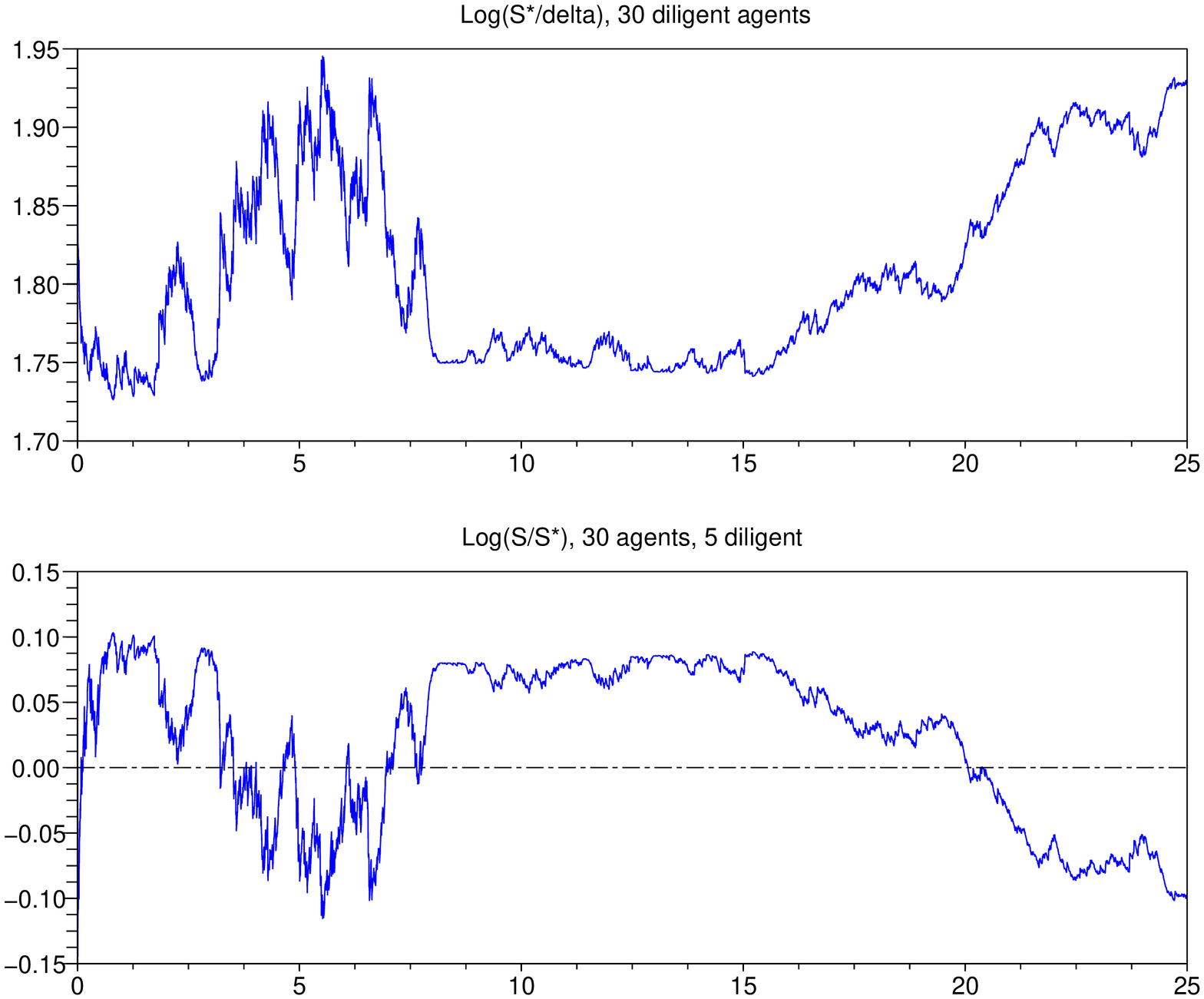}
\end{center}
\caption{
}
\label{Fig5_25Y_30A_5K}
\end{figure}

\begin{figure}[H]
\begin{center}
\includegraphics[height=16.5cm,width=14.5cm,angle=270]
{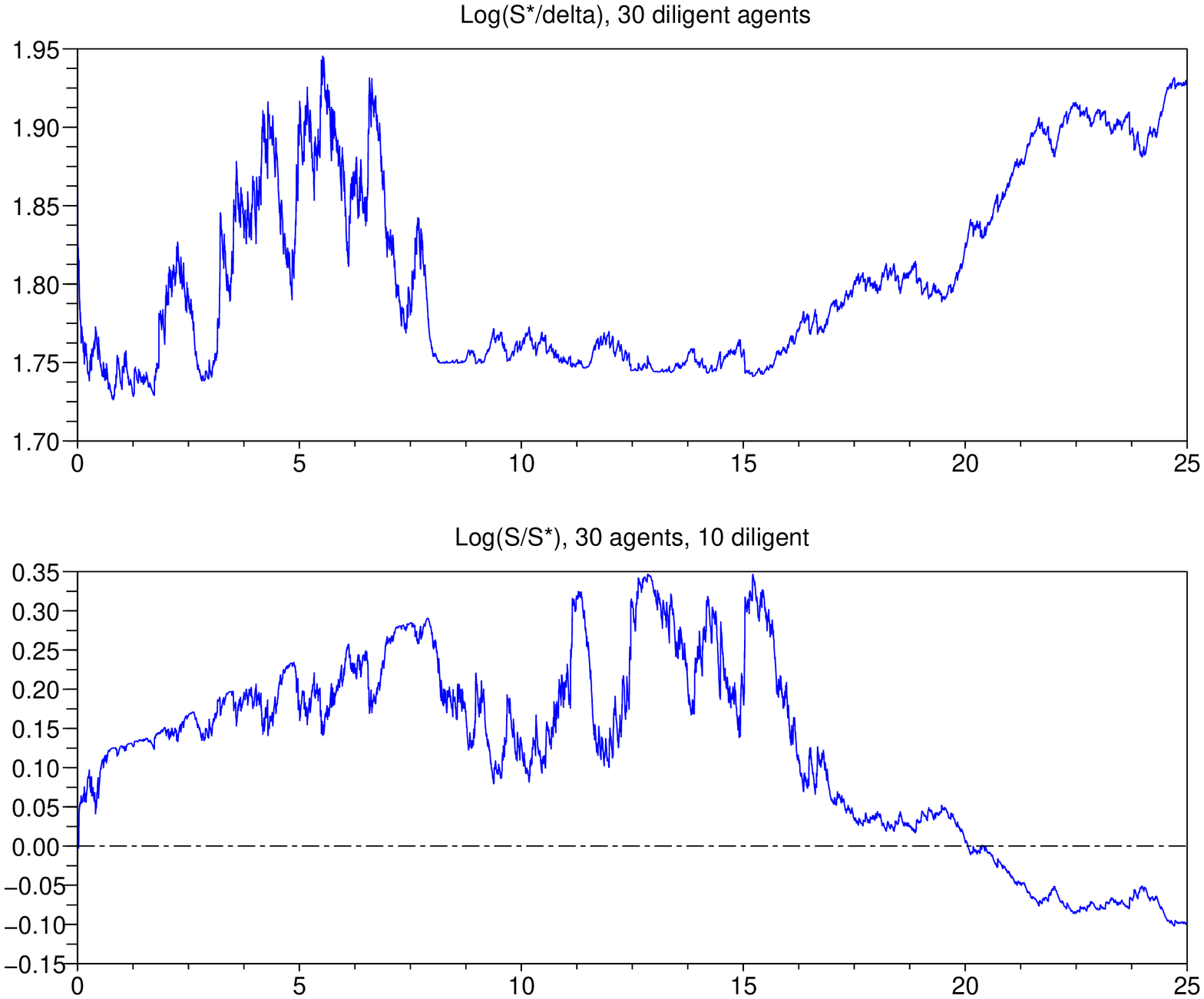}
\end{center}
\caption{
}
\label{Fig5_25Y_30A_10K}
\end{figure}

\begin{figure}[H]
\begin{center}
\includegraphics[height=16.5cm,width=14.5cm,angle=270]
{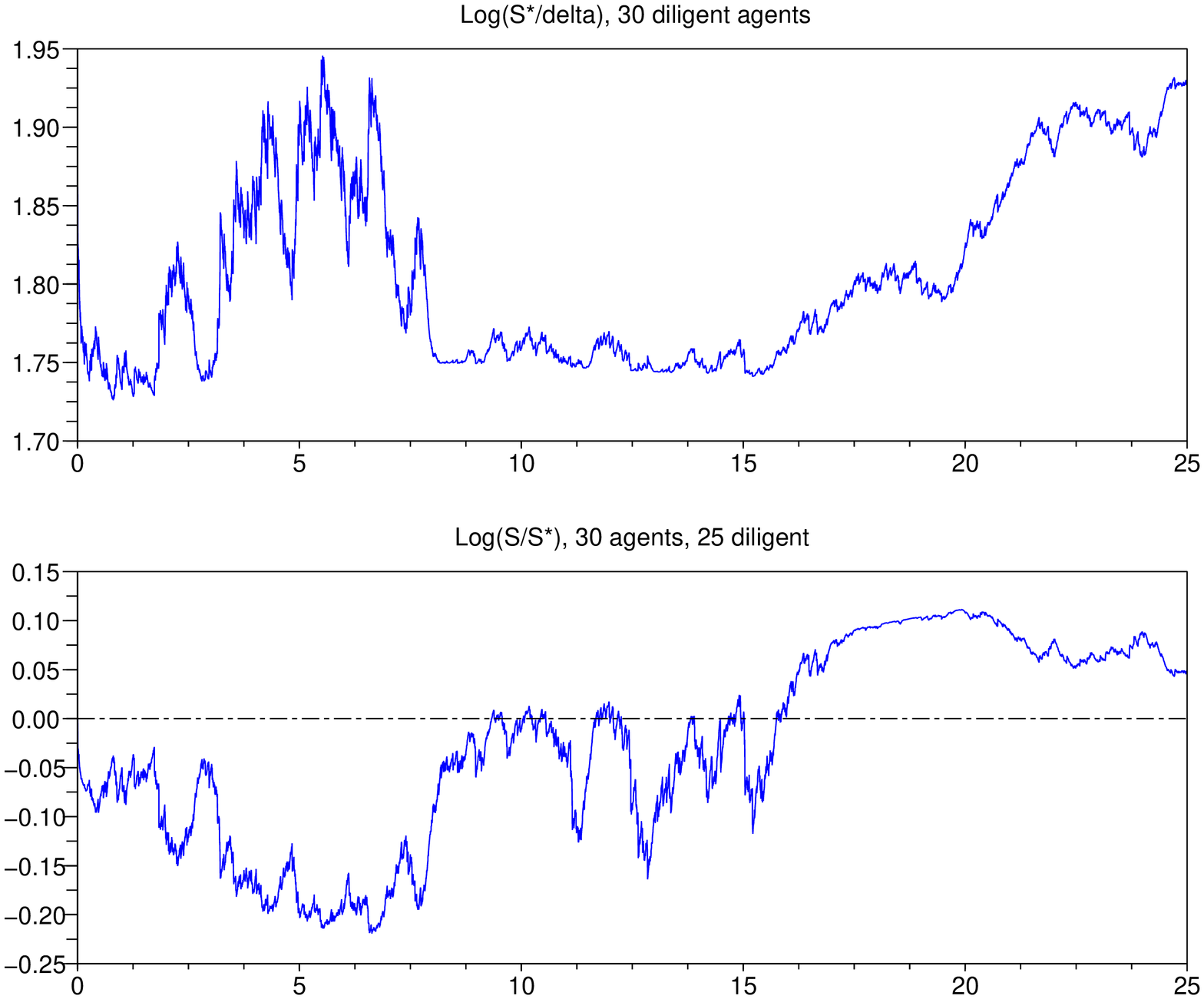}
\end{center}
\caption{
}
\label{Fig5_25Y_30A_25K}
\end{figure}

\begin{figure}[H]
\begin{center}
\includegraphics[height=16.5cm,width=14.5cm,angle=270]
{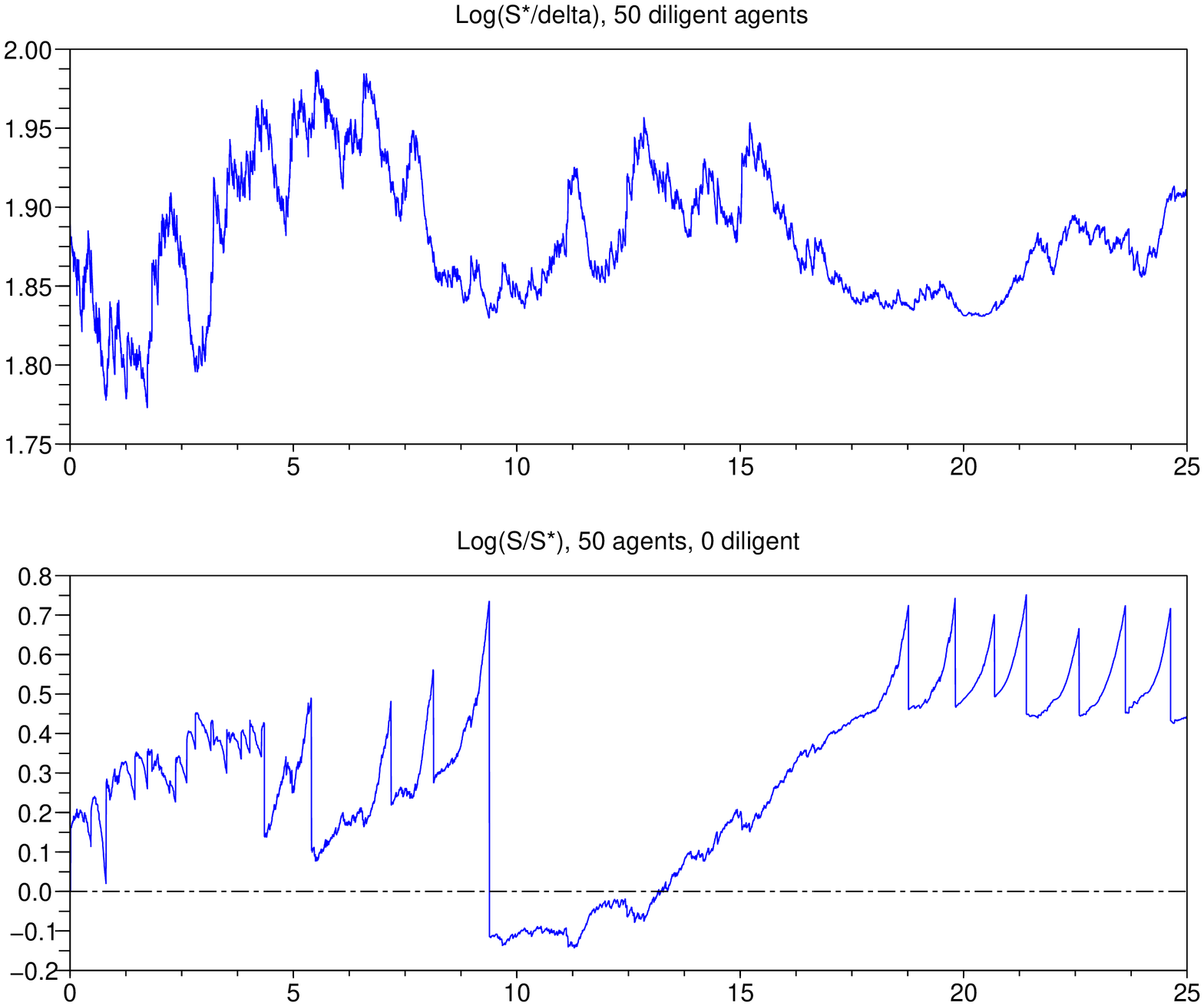}
\end{center}
\caption{
}
\label{Fig5_25Y_50A_0K}
\end{figure}

\begin{figure}[H]
\begin{center}
\includegraphics[height=16.5cm,width=14.5cm,angle=270]
{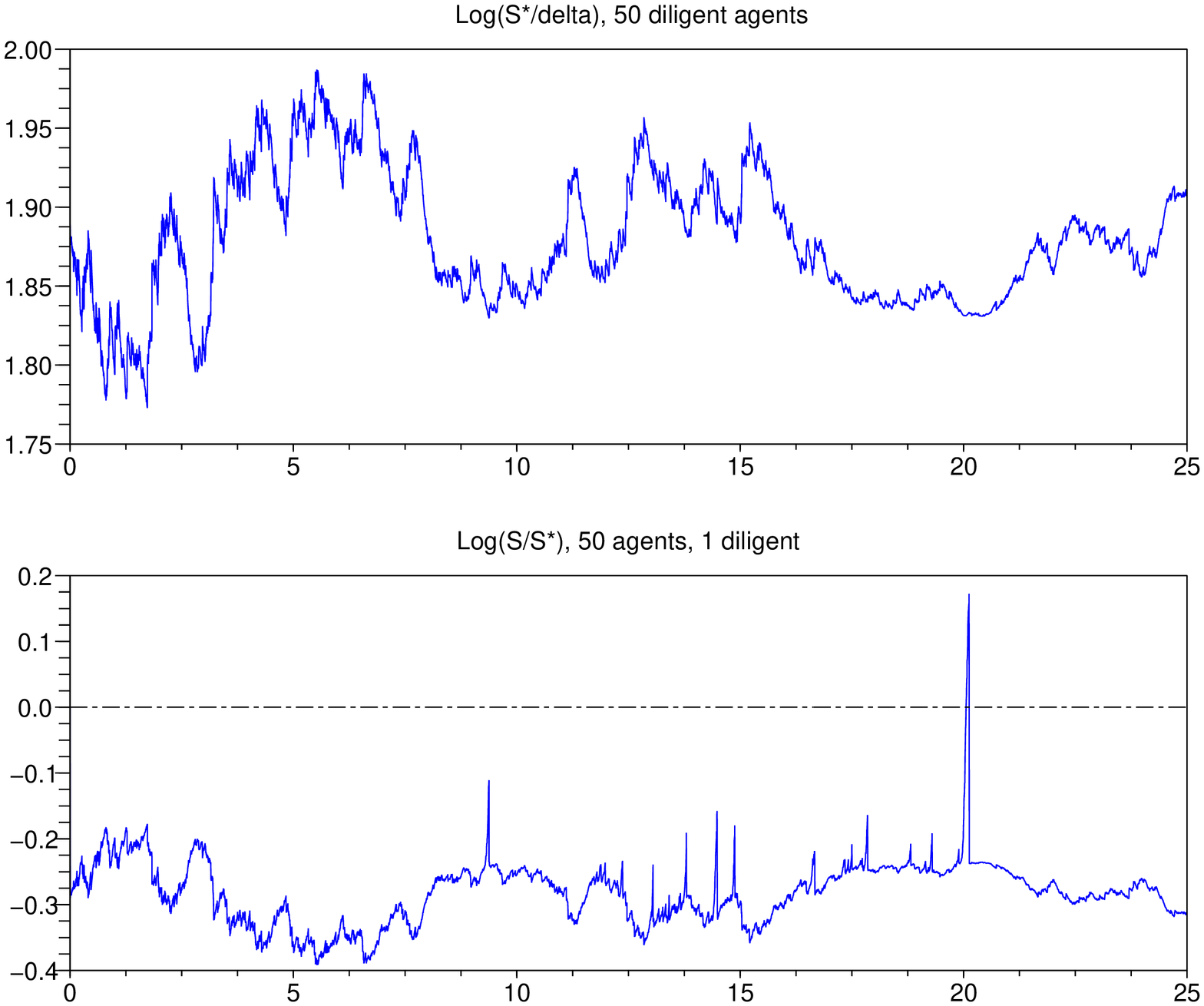}
\end{center}
\caption{
}
\label{Fig5_25Y_50A_1K}
\end{figure}

\begin{figure}[H]
\begin{center}
\includegraphics[height=16.5cm,width=14.5cm,angle=270]
{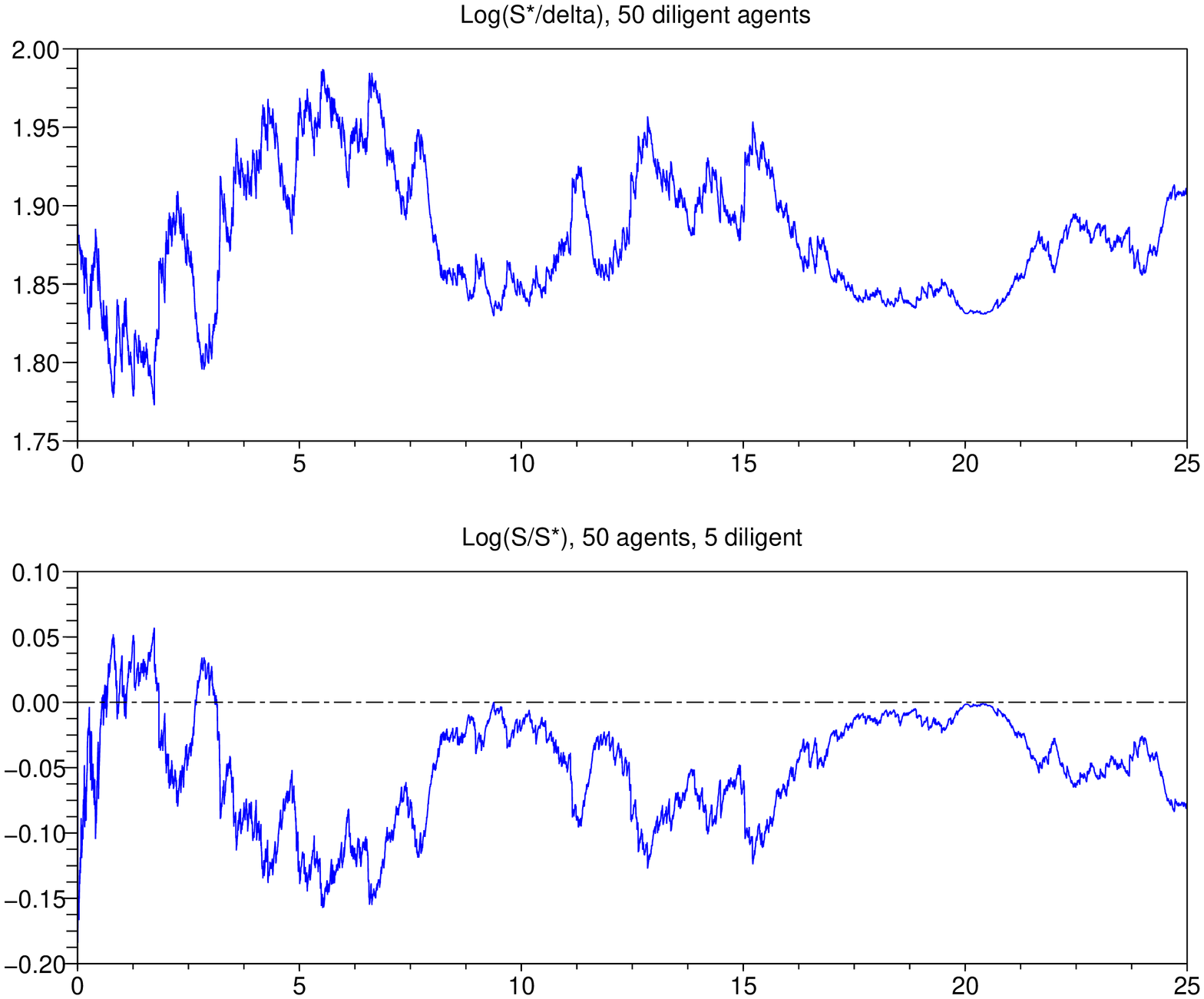}
\end{center}
\caption{
}
\label{Fig5_25Y_50A_5K}
\end{figure}

\begin{figure}[H]
\begin{center}
\includegraphics[height=16.5cm,width=14.5cm,angle=270]
{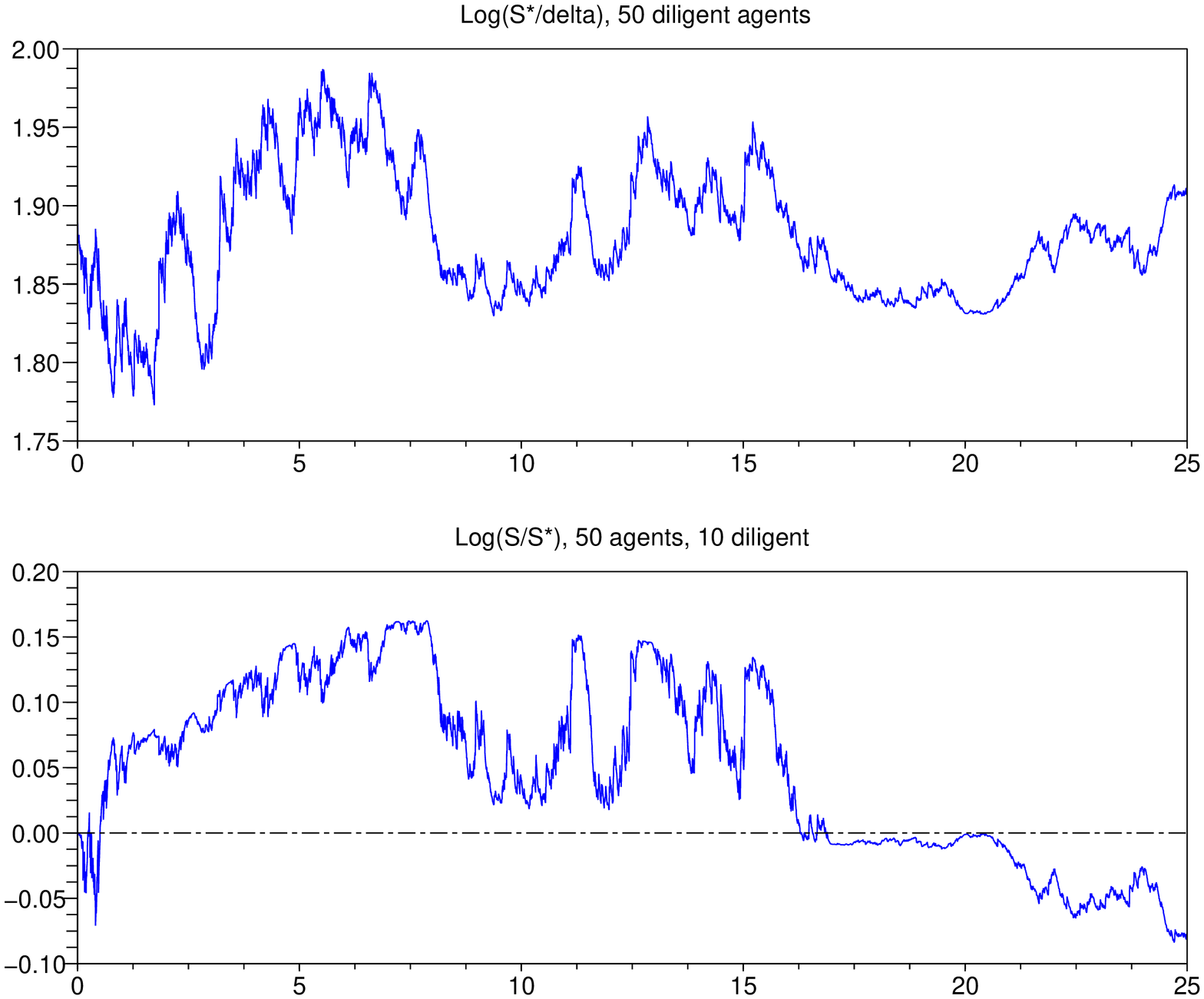}
\end{center}
\caption{
}
\label{Fig5_25Y_50A_10K}
\end{figure}

\begin{figure}[H]
\begin{center}
\includegraphics[height=16.5cm,width=14.5cm,angle=270]
{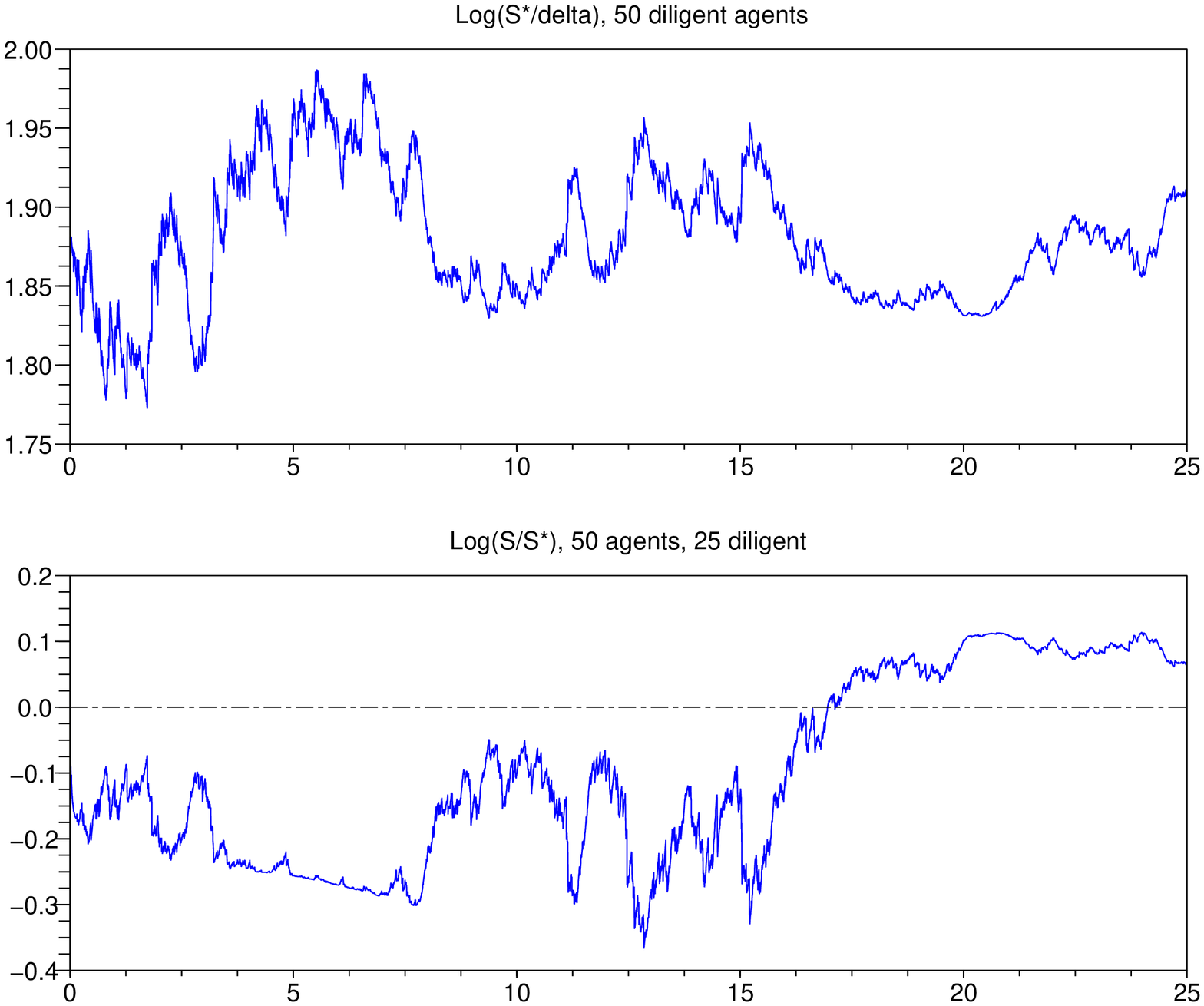}
\end{center}
\caption{
}
\label{Fig5_25Y_50A_25K}
\end{figure}

\begin{figure}[H]
\begin{center}
\includegraphics[height=16.5cm,width=14.5cm,angle=270]
{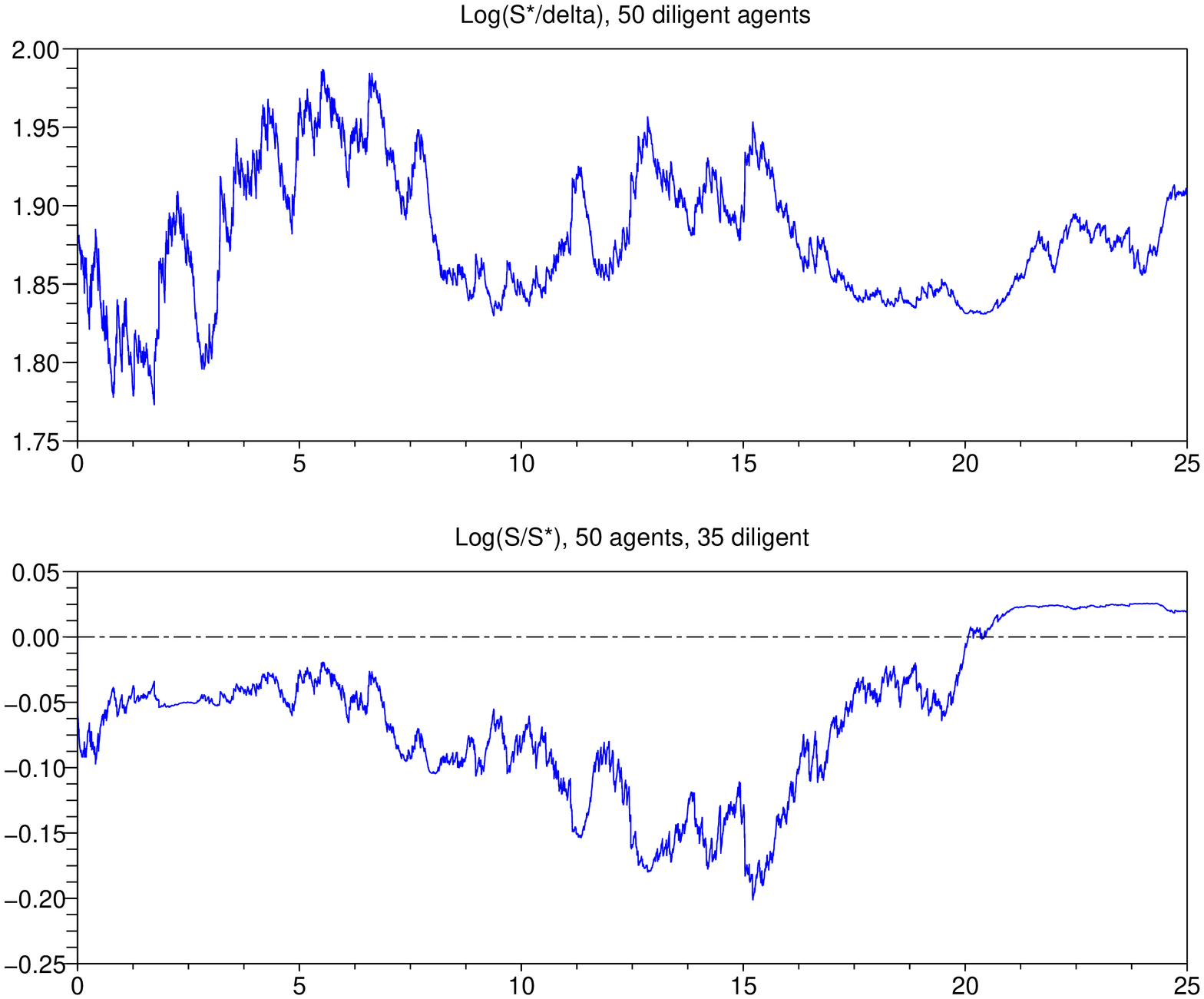}
\end{center}
\caption{
}
\label{Fig5_25Y_50A_35K}
\end{figure}
\section{Proof of Theorem \ref{thmC}}\label{PIDB}
There are several steps to the proof.
\begin{enumerate}
\item[(i)] If we write $\bl^j_t \equiv U'_j (t, \bc^j_t)$, the first thing to
 prove is that  for $0 \leq t <T$
\be
\bl^j_t \bS_t = E\left[\bl^{j}_{t+1} (\bS_{t+1}  + \delta_{t+1}) \ \Big| 
\ \bar{\cal{F}}^j_t\right] .
\label{ce5}
\ee
Consider a (small) perturbation $\bt^j \mapsto \bt^j +\eta$ of the 
portfolio process, with corresponding change $\bc^j \mapsto  
c=\bc^j + \epsilon$ to the consumption process, where (see\eqref{wealthC}) 
\[
\epsilon_t = \eta_t (\bS_t + \delta_t) - \eta_{t+1} \bS_t.
\]
Since $U$ satisfies the Inada condition, $\bc^j$ must be strictly positive 
and  so for small enough $\eta$ the process $c$ will be strictly positive. 
 To leading order the change in agent $j$'s objective is
\bea
E\Big[\sum^T_{t=0} U'_j (t, \bc^j_t)\epsilon_t \Bigr] & = &
           E\Big[\sum^T_{t=0} \bl^j_t \{ \eta_t (\bS_t + \delta_t) - \eta_{t+1} \bS_t\}\Big]\nonumber
 \\
&=& E\Big[\sum^T_{t=1} \eta_t (\bl^j_t (\bS_t + \delta_t) - \bl^j_{t-1} 
                   \bS_{t-1})\Big]\nonumber
\\
& = & E\Big[\sum^T_{t=1} \eta_t E(\bl^j_t (\bS_t + \delta_t) 
- \bl^j_{t-1}  \bS_{t-1}\,  \big| \, \bar{\sF}^j_{t-1})\Big], 
\label{ce6}
\eea
using the facts that $\bS_T=0$, $\eta_0=0$ (since $\theta^j_0=y^j$ is fixed),
 and that the portfolio perturbation must be $\bar{\sF}^j$-previsible.  This
  leading-order change in objective must be $0$, since $(\bt^j,\bc^j)$ was
   optimal; since $\eta$ is arbitrary, inspection of \eqref{ce6} gives \eqref{ce5}.

\item[(ii)] Since $\bc^j$  is $\bar{\sF}^j$-adapted and $\bt^j$ is
 $\bar{\sF}^j$-previsible, we have
\begin{eqnarray*}
\bar{\sF}^j_t &\equiv& \sigma (X_u, \bS_u, z^j_u \ : \ u \leq t) 
\\
& = & \sigma (X_u, \bS_u, \bt^j_{u+1}, \bc^j_u, z^j_u \ : \ u \leq t)
 \\
& \supseteq & \sigma (X_u,\bS_u, \bt^j_{u+1}, \bc^j_u \ : \ u \leq t) 
\equiv {\sF}^j_t,
\\
\end{eqnarray*}
say.  Since $\bl^j_t$ and $\bS_t$ are measurable with respect
 to ${\sF}^j_{t}$, we can refine \eqref{ce5} to 
\be
\bl^j_t \bS_t = E\Big[ \bl^j_{t+1} (\bS_{t+1} + \delta_{t+1})
 \ \Big| \ {\sF}^j_t \Big].
 \label{ce7}
\ee
\item[(iii)] We now take a regular conditional distribution $\kappa^j$ for
 $\bS$ given $(X,\bt^j,\bc^j)$ - see II.89 in \cite{RogersWilliams}. 
  To build the sample
  space $\tOmeg$ on which the DB equilibrium will be constructed, the first
   step is to take $\Omega_0$ to be the path space of $(X,\Theta, C)$, which
    is isomorphic to $\R^{(d+2J)(T+1)}$. This gets its Borel $\sigma$-field
     $\sB$, and canonical filtration; we endow it with a reference probability
      measure $P^*$ which is the law  of $(X,\bT,\bC)$.  Next we expand 
      the sample space to $\tOmeg \equiv \Omega_0 \times \R^{T+1}$,
       and we write $(\tX, \tThe, \tC)$ for the processes defined on $\tOmeg$ by
\begin{eqnarray*}
\tX (\tomeg) & = & X(\omega)\\
\tThe (\tomeg) & = & \Theta (\omega)\\
\tC (\tomeg) & = & C(\omega),\\
\end{eqnarray*}
where $\tomeg = (\omega, s)$, $\omega \in \Omega_0$, 
$s=(s_0,\dots, s_T) \in \R^{T+1}$.  We define a process $\tS$ by
\[
\tS_t (\tomeg) = s_t
\]
when $\tomeg = (\omega, (s_0,\dots, s_T))$.  We write
 $\tilde{\sG}_t = \sigma (\tX_u, \tThe_{u+1}, \tC_u, \tS_u : u \leq t)$ for
  the filtration generated by these processes.

Now we specify the probabilities $P^j$ giving the diverse beliefs of the 
agents.  Firstly, select $(\tX, \tthe^j, \tc^j)$ according to the law
 $P^*$ (equivalently, $(\tX, \tthe^j, \tc^j)$ has the same law as $(X, \bt^j,
  \bc^j)$). Then conditional on $(\tX, \tthe^j, \tc^j)$ let the law of $\tS$  
  be $\kappa^j (\tX, \tthe^j, \tc^j; \cdot)$, and let the random variables
   $\tthe^i, \tc^i$, $i\neq j$, be independent subject to the constraints 
   $\sum \tthe^i_t =1$, $\sum\tc^i_t=\delta_t$. [For example, we could take
    exponential variables $V^i$, $\tilde{V}^j$, and define $\tthe^i_t =
     (1-\tthe^j_t)V^i/\sum_{\ell\neq j} V^\ell$, $\tilde{c}^i_t= 
     (\tilde{\delta}_t - \tc^j_t)\tilde{V}^i/\sum_{\ell\neq j} \tilde{V}^\ell$]. 
      This construction has achieved the following properties:
\item[](a) the $P^j$-distribution of $(\tX, \tthe^j, \tc^j, \tS)$ is the same as
 the $P$-distribution of $(X, \tb^j,\bc^j,\bS)$;

\item[](b) $\tthe^j$ is $\tilde{\sG}$-previsible.

\item[](c) 
\[
\tthe^j_t (\tS_t+\tilde{\delta}_t) = \tthe^j_{t+1} \tS_t + \tc^j_t
\]
with $P^j$-probability 1, since this is a statement about the joint law 
of $(\tX, \tthe^j, \tc^j, \tS)$ and it must therefore have the same 
probability as the corresponding statement about $(X, \bar\theta^j, 
\bar c^j,\bar S)$;

\item[](d) Similarly,
\[
\sum_J \tthe^i_t =1,\quad \sum_J \tc^i_t = \tilde{\delta}_t
\]
with $P^*$-probability 1, and with $P^k$-probability 1 for each 
$k$.
\item[(iv)] Now we define the filtration $\tilde{\sF}^j_t = \sigma 
(\tX_u, \tS_u, \tthe^j_{u+1}, \tc^j_u  \ : \ u \leq t)$, and the processes 
$\tl^j_t \equiv U'_j (t,\tc^j_t)$.  Hence we observe the analogue
\be
\tl^j_t \tS_t =E^j \Big[\tl^j_{t+1} (\tS_{t+1} +\tilde{\delta}_{t+1}) 
\Big| \tilde{\sF}^j_t\Big]
\label{ce8}
\ee
of \eqref{ce7} must hold, because the conditional expedition is determined by 
the joint law of the conditional and conditioning variables, which
 (in view of (a)) is the same as the joint law of the corresponding
  variable in the PI equilibrium for which \eqref{ce7} holds.

\item[(v)]  The next step is to argue that \eqref{ce7} holds when we condition on
 the larger $\sigma$-field $\tilde{\sG}_t$:
\be
\tl^j_t \tS_t =E^j \Big[\tl^j_{t+1} (\tS_{t+1}+\delta_{t+1})\Big| 
\tilde{\sG}_t \Big].
\label{ce9}
\ee
But $\tilde{\sG}_t = \tilde{\sF}^j_t \ \vee \ {\sA}^j_t$, where 
${\sA}^j_t = \sigma (\tc^i_u, \tthe^i_{u+1} \ : \ u \leq t, \ i \neq j)$
 is independent of $\tilde{\sF}^j_t$ so we may apply Proposition \ref{prop3}
  to deduce this result.

\item[(vi)] The final step is to verify the optimality property (v)  in the
 definition of  a DB equilibrium.  Suppose for this that $(\theta_t, c_t)$ is 
 any possible investment-consumption pair for agent $j$ (so
  $\theta_0=y^j$, $\theta$ is $\tilde{\sG}$-previsible, $c$ is 
  $\tilde{\sG}$-adapted, $\theta_t(\tS_t + \delta_t)=\theta_{t+1} \tS_t+c_t$
   for all $t$) and consider the objective
\begin{eqnarray*}
E^j\sum^T_{t=0} U_j (t,c_t) & \leq & E^j \sum^T_{t=0} \Big[U_j (t, \tc^j_t) 
    + \tl^j_t (c_t-\tc^j_t)\Big]
    \\ 
& = & E^j \sum^T_{t=0} \Big[U_j (t, \tc^j_t) + \tl^j_t \Big\{(\theta_t-\tthe^j_t)
(\tS_t+\tilde{\delta}_t) - (\theta_{t+1}-\tthe^j_{t+1})\tS_t\Big\}\Big] 
\\
& = & E^j \sum^T_{t=0} U_j (t, \tc^j_t) + E \sum^T_{t=1}
 (\theta_t-\tthe^j_t)\Big\{ \tl^j_t (\tS_t + \tilde{\delta}_t) - \tl^j_{t-1} \tS_
{t_1}\Big\}
 \\
& = & E^j \sum^T_{t=0} U_j (t, \tc^j_t)\\
\end{eqnarray*}
using (respectively)  concavity of $U_j$, the wealth equation, the fact that 
$\theta_0 = \tthe^j_0=y^j$ and $\theta_{T+1} = 0 = \tthe^j_{T+1}$, 
and \eqref{ce8} together with $\tilde{\sG}$-previsibility of $\theta$, $\tthe^j$. 

\hfill$\square$
\end{enumerate}

\medskip
\noindent
We give here a simple and intuitive result that was needed in the 
proof of Theorem \ref{thmC}.
\medskip
\begin{proposition}\label{prop3}
If $X$ is an integrable random variable, if $\sG$ and $\sA$ are two
sub-$\sigma$-fields of $\sF$ such that $\sA$ is independent of 
$X$ and $\sG$, then
\begin{equation}
  E[X \vert \sG] = E[X\vert \sG\vee\sA] \qquad \hbox{\rm a.s.}
  \label{KeyCE}
\end{equation}
\end{proposition}

\medskip
\noindent{\sc Proof of Proposition \ref{prop3}.} Consider the collection 
\[
\sC \equiv \{ F \in \sF: \int_F \, E[X|\sG] \; dP = \int_F \, X \; dP\}.
\]
This collection is a $d$-system (see \cite{RogersWilliams} Chapter II.1 for 
definitions and basic results). From the definition of conditional 
expectation, $\sG \subseteq \sC$. Now take any $G \in \sG$, 
$A\in\sA$ and calculate
\begin{eqnarray*}
\int_{A \cap G} \, E[X\vert \sG] \; dP &=& \int I_A E[XI_G\vert \sG] \; dP
\\
&=& P(A) E[X:G]  \qquad(\hbox{\rm $\sA$ is independent of  $\sG$})
\\
&=& \int_{A\cap G} X \; dP  
          \qquad(\hbox{\rm $\sA$ is independent of  $\sG$, $X$}).
\end{eqnarray*}
Thus $\sC$ contains the $\pi$-system $\sI$ consisting of all intersections
of the form $A\cap G$, $A\in\sA$, $G\in\sG$, and by Lemma II.1.8
of \cite{RogersWilliams}, the $d$-system $d(\sI)$ generated by $\sI$ equals
the $\sigma$-field generated by $\sI$, which is $\sG \vee \sA$. But
$\sC$ is a $d$-system, and so contains $\sG\vee\sA$.  This establishes
the result.
\hfill$\square$

\medskip
\medskip
\noindent
{\sc Remark.} Intuitively, it seems plausible that we should not need 
$\sA$ to be independent of $\sG$ for this result to hold; after all, what
we are adding to the $\sigma$-field is independent of the random variable.
But this is not true. Take the example of two independent $B(1,\half)$ random
variables $X$ and $Y$, and let $\sA = \sigma(Y)$, $\sG = 
\sigma(Z)$ where $Z = (X+Y)\mod(2)$. Thus $Z$ is 1 if exactly one of 
$X$, $Y$ is 1, zero otherwise. It is not hard to see that $\sG$ is
independent of $X$, and yet $\sA \vee \sG = \sF$, so the 
equality \eqref{KeyCE} fails.

\pagebreak
\bibliography{MAINbib}
\bibliographystyle{acm}

\end{document}